\documentclass[fleqn,usenatbib]{mnras}

\usepackage{newtxtext,newtxmath}

\usepackage[T1]{fontenc}
\usepackage{xcolor}

\DeclareRobustCommand{\VAN}[3]{#2}
\let\VANthebibliography\thebibliography
\def\thebibliography{\DeclareRobustCommand{\VAN}[3]{##3}\VANthebibliography}

\usepackage{graphicx}	
\usepackage{amsmath}	
\usepackage{caption}  

\usepackage{booktabs}
\usepackage{longtable}
\usepackage{url}

\title[JCMT--Venus: Daily Monitoring of HDO and SO$_2$ Absorption in Venus' Mesosphere]{JCMT--Venus: Daily Monitoring of HDO and SO$_2$ Absorption in Venus' Mesosphere}

\author[Tang et al.]{
\scalebox{0.948}{Wei Tang,$^{1}$\thanks{E-mail: wei.tang16@imperial.ac.uk}
David L. Clements,$^{1}$
Jane Greaves,$^{2}$
Malcolm J. Currie,$^{3}$
Supachai Awiphan,$^{4}$
Helen Jane Fraser,$^{5}$}
\newauthor
\scalebox{0.948}{Masataka Imai,$^{6}$
Ing-Guey Jiang,$^{7}$
Xiaohu Li,$^{8,9}$
Kim Miskovetz,$^{10}$
David A Naylor,$^{11}$}
\newauthor
\scalebox{0.948}{Bannawit Pimpanuwat,$^{12}$
Anita M. S. Richards,$^{13}$
Gerald Schieven,$^{14}$
Stephen Serjeant,$^{15}$
and Sirinrat Sithajan$^{4}$}
\\
$^{1}$Department of Physics, Imperial College London, Prince Consort Road, London SW7 2AZ, UK\\
$^{2}$Cardiff Hub for Astrophysics Research and Technology (CHART), School of Physics and Astronomy, Cardiff University,\\
\phantom{$^{2}$}Queen's Buildings North Building, 5 The Parade, Newport Road, Cardiff, CF24 3AA, UK\\
$^{3}$RAL Space, UKRI STFC Rutherford Appleton Laboratory, Chilton, Didcot, Oxfordshire OX11 0QX, UK\\
$^{4}$National Astronomical Research Institute of Thailand (Public Organization), 260 Moo 4, Donkaew, Mae Rim, Chiang Mai 50180, Thailand\\
$^{5}$School of Physical Sciences, Faculty of Science, Technology, Engineering and Maths, The Open University, Walton Hall, Milton Keynes, MK7 6AA, UK\\
$^{6}$Department of Applied Physics, School of Advanced Engineering, Kogakuin University 4-704,\\
\phantom{$^{6}$}Hachioji Campus, Kogakuin University, 2665-1 Nakano, Hachioji, Tokyo 192-0015, JAPAN\\
$^{7}$Department of Physics and Institute of Astronomy, National Tsing-Hua University, Hsinchu 30013, Taiwan\\
$^{8}$State Key Laboratory of Radio Astronomy and Technology, Xinjiang Astronomical Observatory, CAS, 150 Science 1-Street, Urumqi, Xinjiang, 830011, China\\
$^{9}$Xinjiang Key Laboratory of Radio Astrophysics, 150 Science 1-Street, Urumqi, Xinjiang 830011, China\\
$^{10}$JCMT, East Asian Observatory, 660 N. A'ohoku Place, University Park, Hilo, Hawaii 96720, USA\\
$^{11}$Department of Physics and Astronomy, University of Lethbridge, AB, Canada\\
$^{12}$Radio Astrophysics Research Group, National Astronomical Research Institute of Thailand, 260 Moo 4 Donkaew, Maerim, Chiang Mai 50180, Thailand\\
$^{13}$Jodrell Bank Centre for Astrophysics, Department of Physics and Astronomy, The University of Manchester, UK\\
$^{14}$National Research Council -- Herzberg Astronomy \& Astrophysics Research Centre, 5071 West Saanich Road, Victoria, British Columbia V9E 2E7, Canada\\
$^{15}$School of Physical Sciences, The Open University, Walton Hall, Milton Keynes, MK7 6AA, UK
}

\date{Accepted 2026 November 12}

\pubyear{2026}

\begin{document}
\label{firstpage}
\pagerange{\pageref{firstpage}--\pageref{lastpage}}
\maketitle

\begin{abstract}
We present three multi-week, disc-integrated James Clerk Maxwell Telescope (JCMT) campaigns monitoring HDO and SO$_2$ absorption in Venus' upper mesosphere at altitudes of approximately 80--110~km. Obtained in February 2022, July 2023, and August--September 2023, these observations provide dense temporal sampling and complementary observing geometries, enabling the disentanglement of short-term, long-term, and local solar time (LST) variability, which has been challenging in previous studies. We have four main findings. First, SO$_2$ in the upper mesosphere, being largely absent in February 2022 but present in both 2023 campaigns, may follow the same long-term trend observed at cloud tops. Second, long-term variability in H$_2$O is identified here for the first time and is interpreted as an anticorrelated response to the long-term change in SO$_2$, likely through H$_2$SO$_4$ formation. Third, H$_2$O is broadly stable within each campaign, although occasional short-term variability is present in the form of a depletion event. Finally, short-term variability is likely present for SO$_2$, but it does not dominate the overall behaviour. {In the absence of strong long-term differences, the daily SO$_2$ abundance evolution is consistent with an asymmetric LST profile peaking near midnight, which is well reproduced by an independent model.} These results are enabled by a dedicated reduction and retrieval pipeline that mitigates the strong standing-wave baseline structure in the spectra and quantifies systematic uncertainties, providing a methodological framework for future studies facing similar challenges.
\end{abstract}




\section{Introduction}

Venus is our closest planetary neighbour and the only terrestrial planet other than Earth in the Solar System possessing a dense atmosphere. It therefore provides a valuable benchmark for understanding atmospheric processes on terrestrial planets {\citep{kane2019venus}}, particularly those with similarly thick atmospheres for which transmission spectroscopy is most effective. Despite its proximity and importance, Venus has received comparatively limited attention since the numerous missions of the last century, leaving many aspects of Venus' atmospheric chemistry and dynamics poorly understood.

One of the poorly understood aspects of Venus' atmosphere is the behaviour of sulphur dioxide (SO$_2$) and water vapour (H$_2$O) {(see \citealt{marcq2018composition} and references therein)}. These species are key components of the photochemical cycle operating in the planet's CO$_2$-rich atmosphere, which is responsible for the formation of the global sulphuric acid (H$_2$SO$_4$) cloud deck (48--70~km) and hazes. In the consensus model, SO$_2$ and CO$_2$ are photodissociated by solar ultraviolet radiation in the mesosphere (60--110~km), where photochemistry dominates, producing atomic oxygen that oxidises SO$_2$ to SO$_3$. This SO$_3$ subsequently reacts with H$_2$O to produce H$_2$SO$_4$ vapour, which condenses into droplets that sediment downwards to form the main cloud deck  \citep{yung1982photochemistry}. At altitudes below 48~km, H$_2$SO$_4$ thermally dissociates, releasing SO$_2$ and H$_2$O, which are then transported upward, likely through mechanisms involving convection and Hadley circulation, thus completing the cycle \citep{krasnopolsky2007chemical}.

Many observed behaviours of these two species, including their vertical distribution and temporal variability, cannot be explained by the consensus model alone \citep{vandaele2017sulfur1,vandaele2017sulfur2,rimmer2021hydroxide}. Temporal variability, which is the focus of this paper, has long been reported for both species in the mesosphere. Near the cloud tops (60--70 km), the variability of both species has been well characterised. SO$_2$ has been monitored using ultraviolet observations from multiple orbiters, including Pioneer Venus, Venus Express, and Akatsuki, spanning 1978--2022 and revealing both decadal long-term variations and short-term changes on the order of a few Earth days \citep{esposito1984sulfur, marcq2020climatology, iwanaka2025sulfur}. H$_2$O has been monitored using Venus Express infrared observations from 2006--2014, which reveal strong short-term variability but no prominent long-term trend \citep{fedorova2016variations, cottini2015water}. These findings are broadly supported by ground-based thermal infrared spectroscopy, which probes both species at similar altitudes with sparse temporal sampling between 2012 and 2025 \citep{encrenaz2025hdo}. These ground-based observations also indicate both decadal long-term trends and short-term variability in SO$_2$, including variations on timescales of hours, while H$_2$O, observed through its proxy HDO, varies moderately on both short and long timescales \citep{encrenaz2016hdo}. A dependence on local solar time (LST) has been identified for SO$_2$, likely driven by semidiurnal thermal tides, whereas no clear LST dependence has been identified for H$_2$O. This LST trend explains part of the observed variability in SO$_2$, although some disagreement remains regarding its pattern. \citet{encrenaz2025hdo} and \citet{marcq2020climatology} report a semidiurnal pattern with maxima near the morning and evening terminators, whereas \citet{iwanaka2025sulfur} instead find a single peak in the afternoon. For variability beyond the LST dependence, several mechanisms have been proposed over different timescales, including solar-cycle modulation, chemical bifurcation, secular variations in the Hadley circulation or convective transport, and episodic volcanic injections \citep{encrenaz2025hdo,marcq2018composition,vandaele2017sulfur1,kopparla2020recharge,marcq2013variations,esposito1984sulfur}. However, no consensus explanation has yet emerged.

The upper mesosphere differs markedly from the cloud-top region in many aspects. Dynamically, it represents a transition from the superrotation at the cloud tops to the subsolar to antisolar (SS–AS) flow above. Chemically, photochemical processes play a stronger role owing to the increased solar flux. This region has been probed by multiple submm/mm observations \citep{sandor2005water,gurwell2007swas,sandor2010sulfur,encrenaz2015submillimeter,piccialli2017mapping}, as well as by solar and stellar occultation measurements onboard Venus Express for SO$_2$ \citep{mahieux2015venus,belyaev2017night}, and by solar occultation measurements (but no stellar occultation) for H$_2$O \citep{bertaux2007warm,fedorova2008hdo,chamberlain2020soir}. Despite numerous observations, the upper mesosphere remains much less well characterised than the cloud-top region. Submm/mm observations provide very sparse temporal sampling, while solar occultation measurements are restricted to the terminators (LST of 6 and 18 h). Although stellar occultation compensates for the limitation, it is available only for SO$_2$ and remains confined to the nightside. In addition, each occultation measurement samples only a limited latitude range.

Short-term variability is commonly reported for SO$_2$ in the upper mesosphere, but its amplitude and timescale remain poorly constrained. The situation for H$_2$O is less clear, with some studies reporting strong short-term variability \citep{sandor2005water,gurwell2007swas,chamberlain2020soir}, while others find little or no clear evidence for it \citep{bertaux2007warm,fedorova2008hdo}. No clear long-term trend has been identified for either species, although \citet{belyaev2017night} suggest the presence of long-term variability in SO$_2$ without further discussion. LST variability has been reported for SO$_2$, but with only limited indications of a diurnal difference \citep{sandor2010sulfur} and higher abundances near midnight than at the terminators \citep{belyaev2017night}, and no well-constrained pattern has yet been established. Therefore, a comprehensive picture of the upper mesosphere is still lacking, and the characterisation of short-term and LST variability, as well as the identification of any long-term trend, remains challenging because at least two of the spatial, short-term temporal, LST, and potential long-term behaviours are always entangled in each individual study. Combining results from different studies is not straightforward, as reported values can differ significantly even over the same periods, likely due to inconsistencies in instrumentation and data reduction. For example, \citet{sandor2005water} report markedly different H$_2$O abundances, with values as low as $\sim$0.1 ppm, compared with $\sim$4.5 ppm from SWAS observations over an almost identical period in December 2002 \citep{gurwell2007swas}.

This highlights the need for consistent, long-duration observations with dense sampling. Our JCMT--Venus project \citep{Clements2024}, a legacy programme comprising three multi-week campaigns to probe the upper mesosphere of Venus through disc-integrated spectroscopy, is well suited to this purpose. The primary focus of the project is phosphine (PH$_3$), a potential biomarker {\citep{sousasilva2020phosphine}}, with the aim of verifying its existence and constraining its origin following its detection by \citet{greaves2021phosphine}. Our spectral bandwidth covers transitions of PH$_3$, HDO, SO$_2$, and other species, thereby enabling their simultaneous monitoring on a daily basis and providing valuable insight into the behaviour of H$_2$O and SO$_2$ in the upper mesosphere.

In this paper, we describe the observational setup in Section~\ref{ob}, the data-processing procedures in Section~\ref{process}, and the daily abundances in Section~\ref{result}. In Section~\ref{discussion}, we discuss the short-term, long-term, and LST-dependent variability of H$_2$O and SO$_2$. We summarise our conclusions in Section 6.

\section{Observations}
\label{ob}

The James Clerk Maxwell Telescope (JCMT) is a 15-metre millimetre/submillimetre telescope located near the summit of Mauna Kea, Hawaii. Our observations use its `Ū`ū receiver, one of the three heterodyne receiver inserts within the Nāmakanui instrument \citep{mizuno2020commissioning}. The `Ū`ū receiver  features dual-polarisation, sideband-separating (2SB) mixers, delivering four intermediate frequencies (IFs) via a local oscillator tunable between 221 and 264.6 GHz. The receiver is coupled to the ACSIS (Auto-Correlation Spectral Imaging System) backend, which offers several bandwidth configurations. We use the hybrid mode, in which each IF input produces two 1000 MHz spectral windows (SPWs) that slightly overlap. Each SPW comprises 2048 channels, yielding a spectral resolution of 0.488 MHz per channel.

The JCMT–Venus project monitors Venus across three observing campaigns: 1–20 February 2022, 6–27 July 2023, and 27 August–24 September 2023. These observing windows were selected to coincide with epochs when Venus is sufficiently large (>36~arcsec) to fill the telescope beam, which has a full width at half power (FWHP) of 18.4~arcsec. This configuration enables sensitivity to the much weaker phosphine signal. Observations during the first and third campaigns were conducted in the morning, and the second campaign took place in the evening. The three campaigns yield 191, 122, and 114 observations, respectively. Each observation consists of four 30-second subscans, which are co-added internally to produce a single integrated scan. The tuning frequency is set to 266.7445\,GHz for the upper sideband (USB) in the first campaign and is later adjusted to 266.8445\,GHz in subsequent ones. This shift moves the SO$_2$ transition away from the spectral boundary, as we initially suspected that its proximity might hinder the detection, but later found that it does not. The observing conditions are constrained by the JCMT Band~4/5 boundary, which requires the 225-GHz atmospheric opacity to be below 0.2. A detailed summary of the three campaigns, including the number of observations, observing times, and daily weather conditions, is provided in \citet{Clements2024}.

 \section{Data Reduction}
\label{process}

\subsection{Pre-processing }

We have developed a JCMT pre-processing tool, built on the \textsc{starlink} software suite \citep{currie2014starlink}, to perform the initial reduction steps. The tool allows the selection and combination of sidebands, polarisations, and SPWs, and can also apply the Venus--Earth velocity correction, transforming each spectrum between the Venus-centric and geocentric frames. 

This study uses data only from the USB, which covers the SO$_2$ $13_{3,11}$--$13_{2,12}$ line at 267.5374~GHz in SPW~1 and the HDO $2_{2,0}$--$3_{1,3}$ line at 266.1611~GHz in SPW~2. The two polarisations are averaged in all campaigns, except for SO$_2$ in Campaign~2, for which only polarisation channel 1 is used because the other shows much stronger ripple in the corresponding SPW. The tool removes the noisy spectral boundaries from each SPW, after which each SPW spectrum is normalised by its continuum level (mean antenna temperature) to correct for continuum variations that affect the apparent line depth. The associated pre-processing parameter choices are detailed by \citet{Clements2024}.

\subsection{Baseline ripples}

Standing-wave artefacts arising from Venus’ exceptional brightness present the main challenge to our analysis. The brightness of Venus causes radiation to reflect within the instrument optics and telescope and cabin environments, with a fraction re-entering the receiver after a delay. When these delayed components are processed by the backend autocorrelator and subsequently Fourier-transformed into the spectral domain, they manifest as sinusoidal baseline ripples in the final spectra, as described by \citet{barnes2005postcorrelation} and \citet{Greaves2024}.

\begin{figure}
    \centering
    \includegraphics[width=\columnwidth]{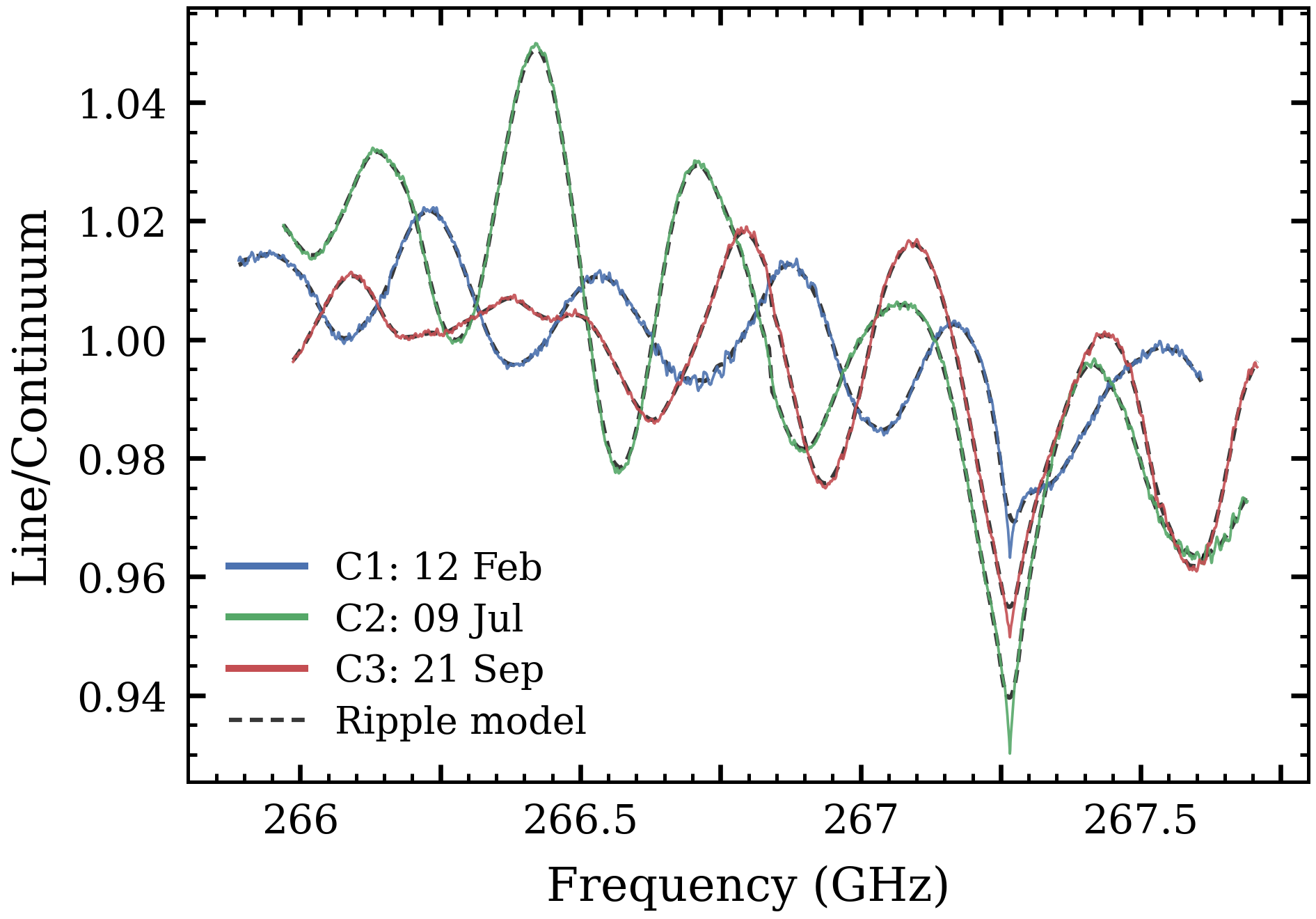}
    \caption{Daily averaged spectra for the dates with the largest number of scans in each of Campaigns~1--3, shown in the geocentric frequency frame. Coloured solid curves show the spectra, while black dashed curves show the corresponding large-scale ripple models extracted using a mean filter.}
    \label{large}
\end{figure}

In principle, the ripple has a predictable period that is inversely proportional to the delay time of the internally reflected signals, and hence to the corresponding reflection path lengths. In our data, however, the ripples arise from multiple reflection paths, producing a complex and irregular pattern that cannot be captured by a simple sinusoidal model. Fig.~\ref{large} shows the dominant large-scale ripple, with a characteristic period of approximately 300 MHz, consistent with reflections between the ambient load and the horn/mixer assembly, which are separated by 0.5 m. This primary ripple is removed from each individual scan by subtracting a mean-filtered ripple model, computed using sliding windows of 64 spectral channels. A median filter was also tested but not adopted, as it over-fits small-scale structure, including genuine spectral features.

\begin{figure}
\includegraphics[width=\columnwidth]{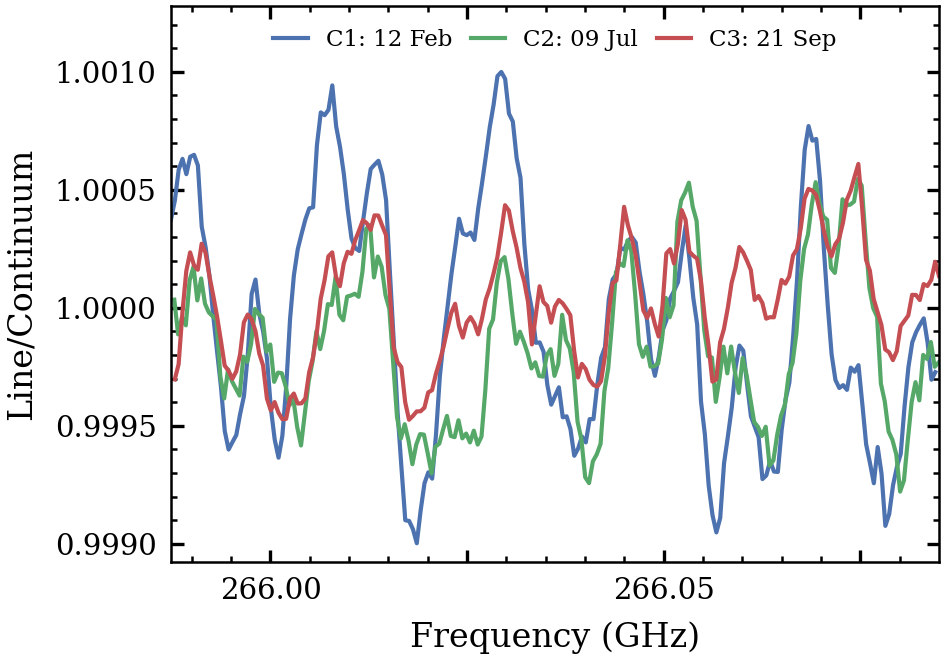}
    \caption{
Small-scale ripples remaining after removal of the mean-filtered large-scale ripple. Coloured curves show daily averaged spectra for the selected dates with the largest number of scans in Campaigns~1, 2, and 3, in the geocentric frequency frame. A randomly selected 200-channel interval is displayed to illustrate the broadly aligned ripple structure across the three campaigns.
}
    \label{small}
\end{figure}

Ripples with smaller amplitudes and shorter periods dominate once the primary ripple is removed, as illustrated in Fig.~\ref{small}. These small-scale ripples correspond to reflected signals that have travelled longer path lengths. Their lower amplitudes are either because reflected signals are scattered, so that longer paths allow a smaller fraction to re-enter the receiver, or because longer paths involve multiple reflections that lead to greater energy losses. Unlike the primary ripple, which generally exhibits a stable and repeating pattern across the bandwidth, these small-scale ripples show greater variability in both period and amplitude. This variability indicates a complex mixture of reflection paths, complicating the identification of their origins. Despite this complexity, the small-scale ripple pattern appears broadly aligned from scan to scan in the geocentric frame, even across different campaigns. This behaviour provides strong evidence that the ripples are of instrumental origin. Nevertheless, the residual scan-to-scan variability remains sufficiently large that it prevents the construction of a universal ripple model applicable to all scans, with deviations large enough to mimic or destroy an absorption feature.

\subsection{Fourier cleaning}
\label{fc}

\begin{figure*}
    \centering
    \includegraphics[width=\textwidth]{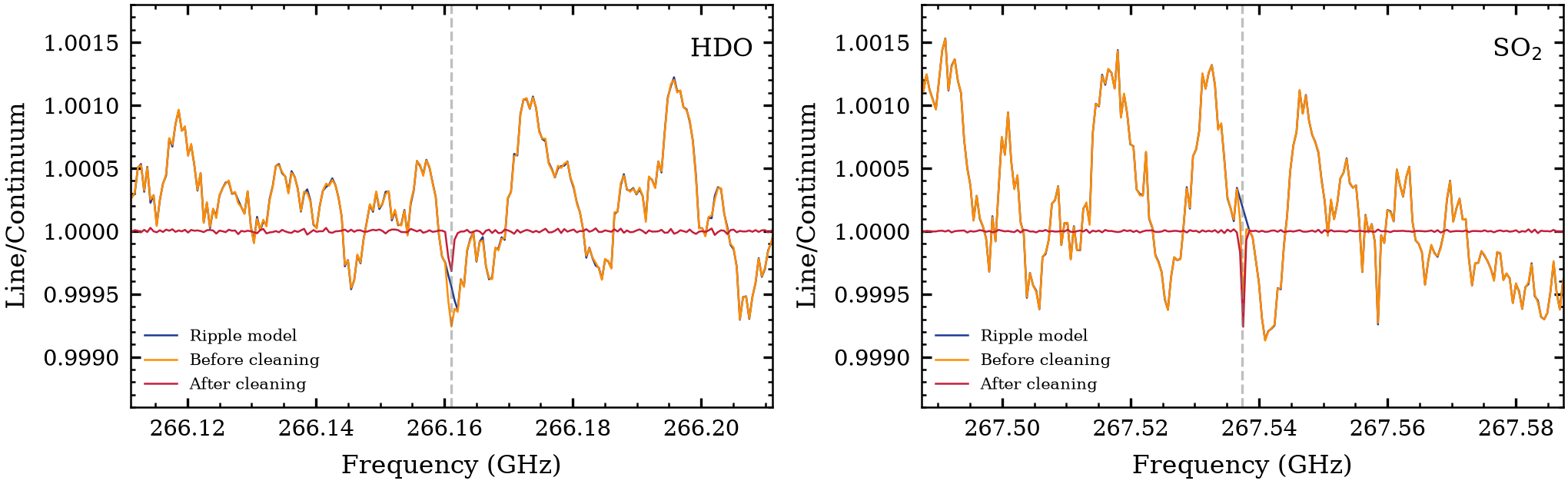}
    \caption{
Example of the Fourier-based removal of small-scale ripples from the 9 July spectrum in Campaign~2. For each species, the orange curve shows the spectrum before Fourier cleaning, the blue curve shows the modelled ripple structure, and the red curve shows the spectrum after subtraction of the ripple model. The left panel shows HDO and the right panel shows SO$_2$. In each panel, the vertical dashed line marks the expected line centre.
}
    \label{fourier}
\end{figure*}

Removal of small-scale ripples requires the construction of a reliable model of the ripple structure, and for this purpose, we adopt a Fourier-based method, relying on the \texttt{FOURIER} command in the \textsc{kappa} package of \textsc{starlink}. The procedure involves averaging scans from the same day, interpolating across a small region centred on the line, Fourier-transforming a wide, truncated section around the line, retaining only the dominant Fourier components associated with the ripple modes while setting all remaining components to zero, and finally applying an inverse transform to produce a ripple-only profile. The ripple model is therefore constructed on a daily basis and subsequently subtracted from each daily averaged spectrum. Fig.~\ref{fourier} shows Fourier cleaning and resulting HDO and SO$_2$ detections for one example day, 9 July, during Campaign 2. This process is repeated, giving line depths, or upper limits, for each day where we have data.

To describe the procedures in more detail, spectra are averaged on a daily basis, as scans obtained within a single day represent the maximum number that can be stacked to reinforce the ripple structure without introducing line dilution. Over a daily timescale, the Venus–Earth relative velocity changes only marginally, and as a result the spectral line, fixed in the Venus-centric frame, and the baseline ripple, fixed in the geocentric frame, remain effectively co-aligned across the scans. The interpolation step linearly connects the two continuum points located at ±2 channels from the line centre. The choice of this four-channel window is motivated by tests performed without interpolation: applying the Fourier cleaning directly still recovers a narrow feature only four channels wide, indicating that the line core within this interval is sufficiently free of ripple contamination to stand out clearly from the ripple structure. Masking these four channels therefore preserves the integrity of the reliable line core while allowing the ripple structure to be modelled as close as possible to the line’s immediate neighbourhood. It is acknowledged that line wings outside this window are incorporated into the ripple model and removed along with it. This has a negligible effect on the abundance determination as discussed in Section~\ref{ar}. A 401-channel section centred on the target line is then selected, providing sufficient baseline information to characterise the ripple behaviour. This section is Fourier transformed, and nearly all Fourier components are retained using a stringent amplitude threshold. Applying the inverse transform to these retained components yields the daily ripple model.

\subsection{Abundance recovery}
\label{ar}
In our reduction, the line wings are removed together with the ripples, leaving a truncated line profile with a manually fixed width of four channels. Because the complete line shape contains vertical information, the loss of the wings prevents a proper retrieval capable of determining the vertical abundance profile. We therefore adopt a forward-modelling approach for the abundance recovery.

In the forward modelling, we simulate spectra on a daily basis by adding modelled lines of various abundances to the daily synthetic baseline, constructed from the large-scale ripple model extracted using the mean filter and the small-scale ripple model obtained from the Fourier cleaning. Before addition, the modelled lines are scaled to match the continuum level of the observed spectrum for that day. Each simulated spectrum is then processed in exactly the same manner as the observed spectrum: it is normalised using the same continuum value and subsequently passed through the mean filter and the Fourier-cleaning procedure. The resulting synthetic line depth shows a strong linear correlation with abundance, as expected given the narrow line profile arising from the low-pressure conditions in our probed altitude range. We therefore determine the best-fitting abundance for the daily observed lines based on their measured depth. As for the issue that the synthetic baseline profile incorporates real line wings, meaning the simulated spectrum would contain two sets of wings, one from the modelled line and one from the real line, we do not think this has any significant influence on the recovered abundance. However many wings are added, they are removed by the subsequent reduction, which eliminates all broad structures.

The line modelling is carried out using the Planetary Spectrum Generator (PSG) \citep{villanueva2018planetary}. The abundance of the target molecule is treated as the only free parameter, and a constant vertical profile is assumed from 0--200 km. All other species employ the default vertical profiles embedded within PSG, which draw on \citet{ehrenreich2012transmission}, \citet{bierson2020chemical}, and \citet{vandaele2020composition}. Line parameters are taken from HITRAN (High-Resolution Transmission Molecular Absorption Database) \citep{gordon2022hitran2020}. The adopted CO$_2$ pressure-broadening coefficients, $\gamma_{\rm CO_2}$, are 0.1398 and 0.1717~cm$^{-1}$~atm$^{-1}$ for the HDO and SO$_2$ lines, respectively.

{{For each observing date, we construct temperature and pressure profiles using the Venus Climate Database (VCD) \citep{lebonnois2010superrotation,lebonnois2016wave,gilli2017thermal,martinez2023variability}, based on the latitude, longitude and LST of the corresponding sub-observer point.}} For conversion between H$_2$O and HDO, we adopt the altitude-dependent D/H profile of \citet{mahieux2024unexpected}, which increases from $\sim$162 to $\sim$1519 times VSMOW (Vienna Standard Mean Ocean Water; $1.5567 \times 10^{-4}$) between $\sim$70 and $\sim$108~km. Outside this altitude range, the profile is set at the corresponding boundary values.

{We further assess the sensitivity of the derived abundances to the assumed temperature and D/H profiles in Appendices~\ref{app:temperature} and~\ref{app:dh}, respectively. We compare VIRA with our daily temperature profiles, and test a constant D/H of $200\times$VSMOW, as adopted by, e.g., \citet{encrenaz2025hdo}, together with the lower and upper envelopes of \citet{mahieux2024unexpected}. Varying D/H approximately rescales the H$_2$O abundances, with the scaling factors varying by less than 6\% between dates within each campaign, apart from the near-zero abundance on 19 September. Relative to the Venus International Reference Atmosphere (VIRA) profile \citep{seiff1985models}, the daily temperature profiles lead to maximum absolute abundance differences of 12.9~ppb for H$_2$O and 2.2~ppb for SO$_2$.}

\section{Results}
\label{result}

\subsection{Altitude sensitivity}
\begin{figure}
    \centering
    \includegraphics[width=\columnwidth]{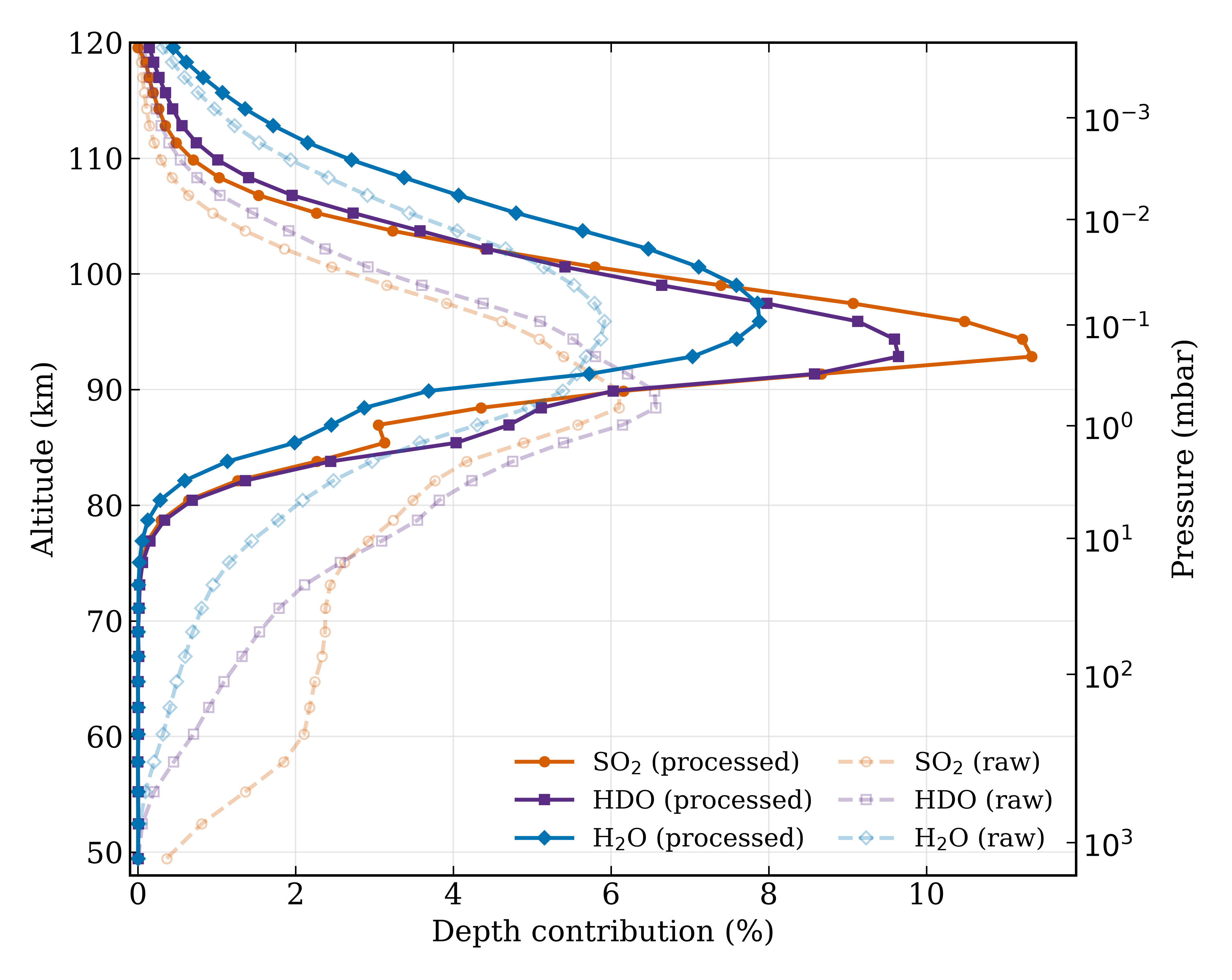}
    \caption{
    Vertical line-depth contribution functions for SO$_2$ (orange), HDO (purple), and H$_2$O (blue). Solid curves show the contribution functions after mean filtering and Fourier cleaning, while dashed curves show the corresponding unprocessed functions. The H$_2$O contribution function is converted from the HDO response using the altitude-dependent D/H profile of \citet{mahieux2024unexpected}. 
    }
    \label{wei}
\end{figure}

We first determine the contribution to the line depth as a function of altitude, which defines the vertical weighting function for our retrieved abundances. This is shown in Fig.~\ref{wei}, calculated by sequentially assigning a constant abundance to each individual atmospheric layer in PSG while setting all other layers to zero, and then evaluating the resulting line depth. We use the mean temperature profile adopted in this study for these calculations. We present both the raw PSG output and the line depth measured after applying our reduction procedure. For H$_2$O, the contribution function is derived from the HDO calculation using the altitude-dependent D/H profile of \citet{mahieux2024unexpected}.

Our observations are sensitive to very similar altitude ranges for SO$_2$ and H$_2$O. After Fourier cleaning, the peak sensitivity shifts slightly towards higher altitudes relative to the raw calculation. The effective sensitivity also becomes more confined, with the dominant contribution arising from approximately 80--110~km, { corresponding to a vertical extent of about 7.5 pressure scale heights.}

\subsection{Daily H$_2$O and SO$_2$ abundances}
\label{daily}

\begin{figure}
    \centering
    \includegraphics[width=\columnwidth]{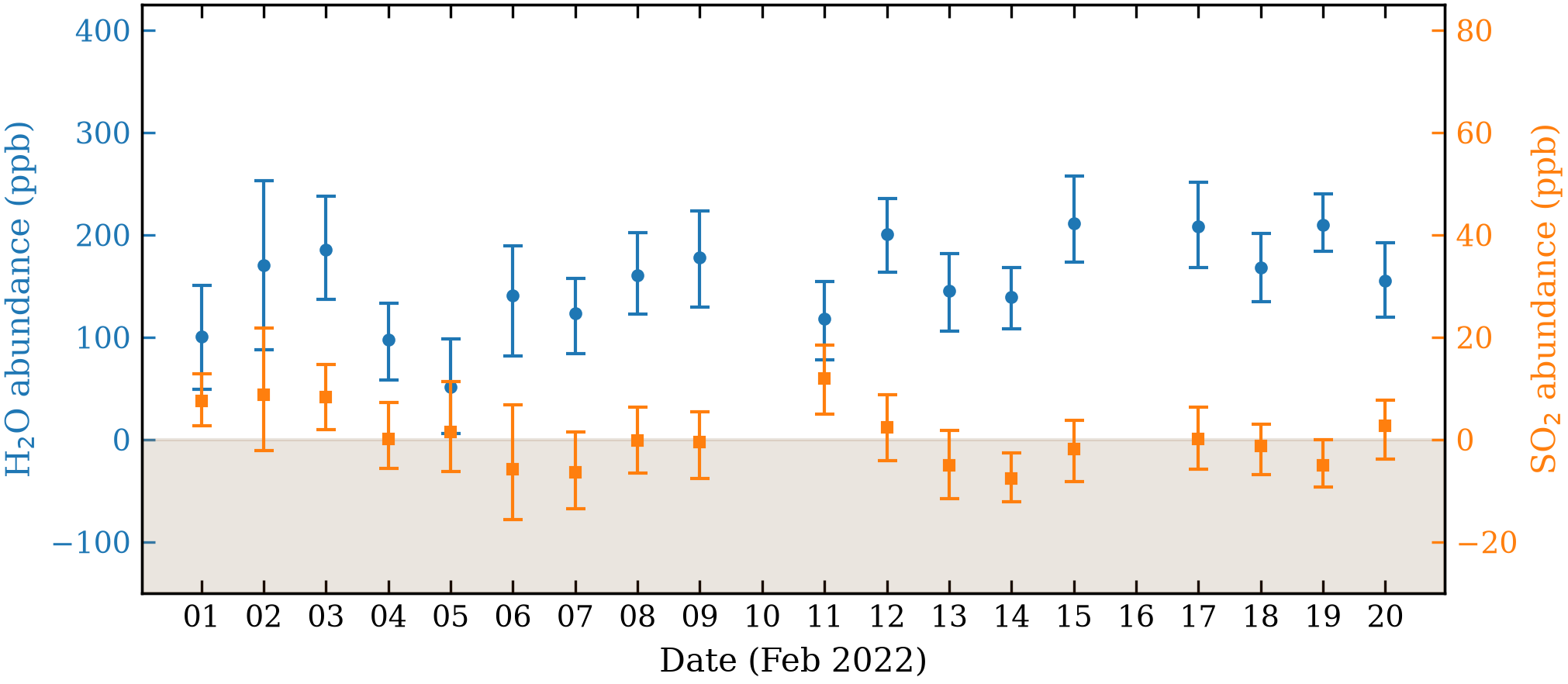}\\[3pt]
    \includegraphics[width=\columnwidth]{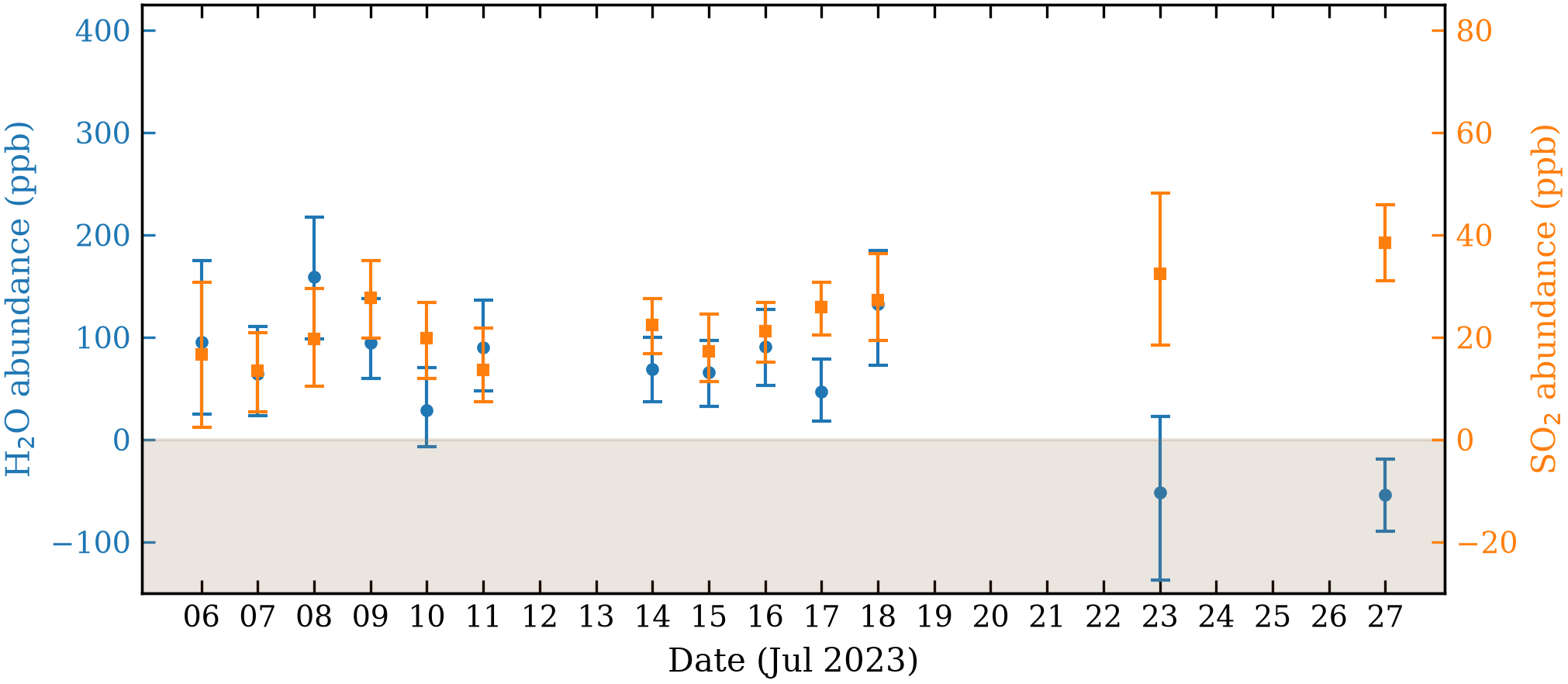}\\[3pt]
    \includegraphics[width=\columnwidth]{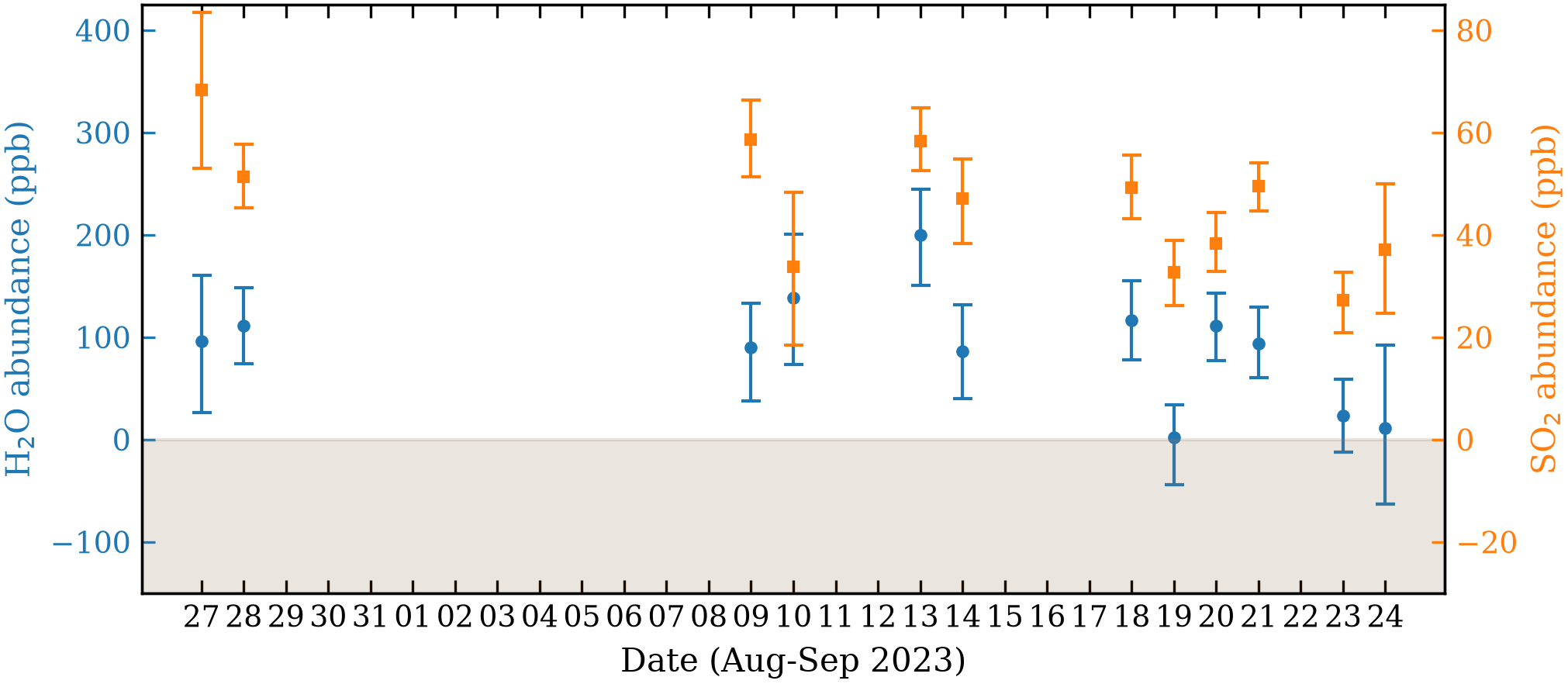}
    \caption{
Daily H$_2$O and SO$_2$ abundances for Campaign~1 (top), Campaign~2 (middle), and Campaign~3 (bottom).
}
    \label{123ab}
\end{figure}

The daily abundances of the two species across the three campaigns, derived using our abundance-recovery method, are shown in Fig.~\ref{123ab}. The abundance values and uncertainties, together with the observing geometry, are summarised in Table~\ref{abundance} in Appendix~\ref{app:daily_data}. The method used to determine the uncertainties is described in Section~\ref{ae}.

SO$_2$ is broadly absent in Campaign~1, and we obtain only tentative detections on a few days. HDO is not detected on two observing days (23 and 27 July in Campaign~2). For all non-detections, we report the pseudo-abundances, which take negative values. These negative values arise because the recovered feature appears in emission rather than absorption, and they are obtained by interpolating the modelled abundances into the negative regime. Although not physically meaningful, these values are statistically informative, providing a quantitative measure of the likelihood that a real signal is present, as well as of apparent line-depth variability beyond that expected from the uncertainties.

The H$_2$O abundances derived here fall within the range reported by \citet{sandor2005water}, whose measurements were based primarily on JCMT data and assumed a D/H of 157, which would yield H$_2$O abundances systematically higher than ours. They obtained values between $0.0 \pm 0.06$ and $3.5 \pm 0.3$~ppm from 17 observations of the 225.8967 and 335.3955~GHz lines acquired between March~1998 and June~2004. Our SO$_2$ abundances are also consistent with the range of $0 \pm 2$ to $76 \pm 9$~ppb reported by \citet{sandor2010sulfur}, based on 21 JCMT observations of the 346.6520~GHz transition obtained between June~2004 and August~2008.

\subsection{Abundance Errors}
\label{ae}
\begin{figure*}
    \centering
\includegraphics[width=\textwidth]{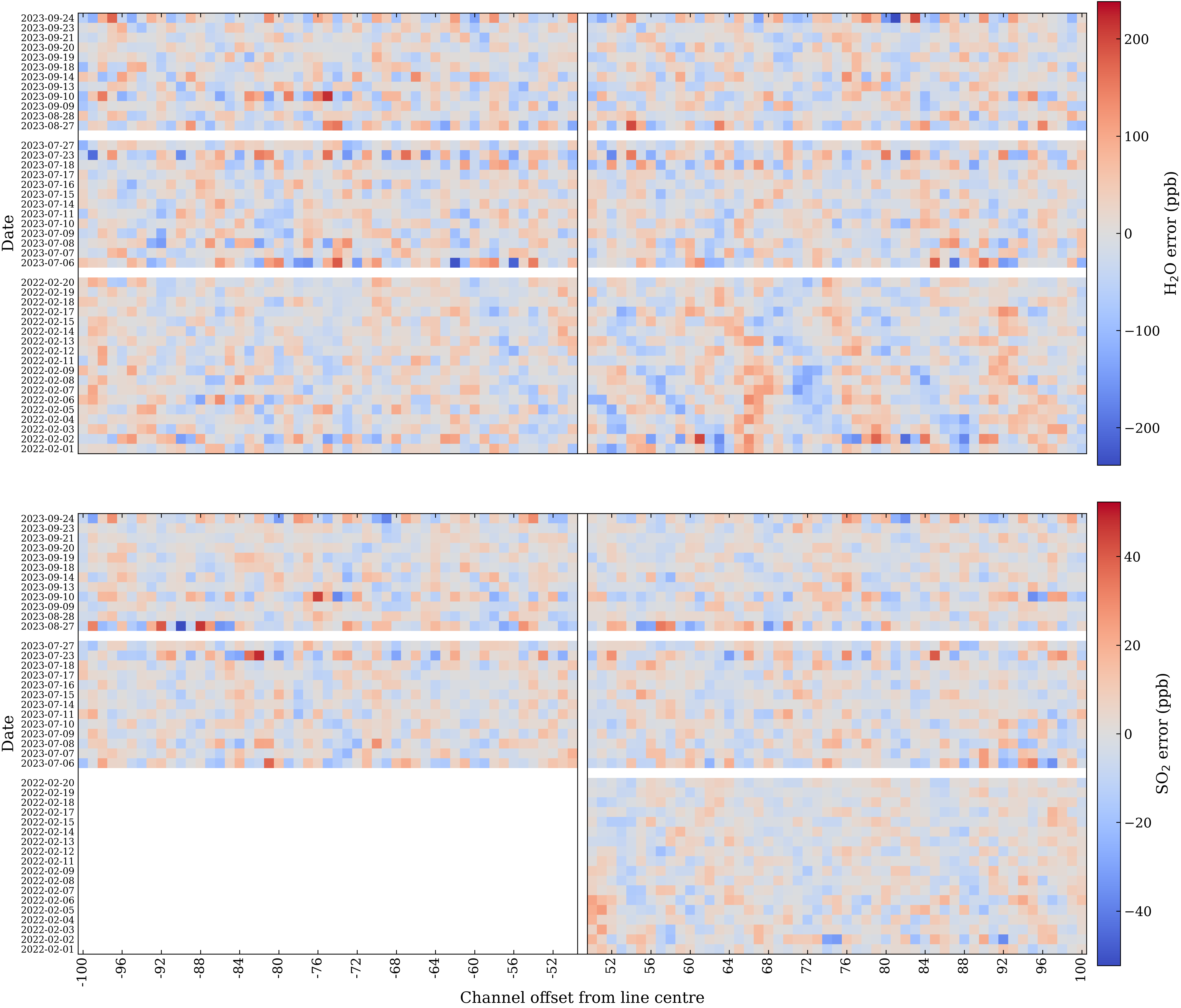}
    \caption{
Abundance error maps from the injection tests. 
Top: H$_2$O error (input minus recovered abundance) as a function of channel offset from the line centre and observing date. 
Bottom: Same as the top panel, but for SO$_2$.
}
    \label{offset_maps}
\end{figure*}

We assess the accuracy of the abundance recovery by injecting modelled spectral lines of known abundance directly into line-free channels of the raw spectra, which is essentially the pure baseline, processing the injected spectra through the same reduction and abundance-recovery pipeline as the real data, and comparing the recovered abundances with the input values. For each daily averaged spectrum, injections are performed with modelled line centres offset by ±50–100 channels from the real line centre, yielding 102 simulated spectra per observing date. The selected injection range reflects the local baseline properties of the target line while ensuring that any contribution from the real line is negligible. Only 51 channels are used for SO$_2$ in Campaign~1 because of the line’s proximity to the spectral edge. The resulting abundance error maps, showing the offsets between input and recovered abundances as a function of injection centre channel and observing date, are presented in Fig.~\ref{offset_maps}. Cells with negative errors indicate cases where the recovered abundance exceeds the input value, while positive cells represent underestimation.

The offsets are typically small, with only a few isolated large values, indicating the reliability of the reduction and abundance-recovery methods. These offsets mainly arise from residual baseline structure overlapping the line core, which is preserved by the interpolation step during Fourier cleaning. Specifically, a ripple peak at the line centre prior to line injection leads to an underestimation of the recovered abundance, whereas a trough results in an overestimation. The magnitude of the offset is determined by the amplitude of the residual baseline structure and is therefore independent of the input abundance. That is to say, the error is independent of the physical abundance.

The error maps are used to determine the uncertainties on the absolute daily abundances presented in Section~\ref{daily}. For each observing date, the corresponding row is extracted from the map, and the negative and positive 1-$\sigma$ uncertainties are defined as the 16th and 84th percentiles of the error distribution, respectively. Larger uncertainties on certain dates arise from the smaller number of scans obtained on those days. This approach to uncertainty estimation is justified by the fact that the error distributions for most dates are consistent with normality according to the Shapiro-Wilk test, with only minor skewness present on a few dates. In this way, both systematic effects introduced by the pipeline and random noise contributions are incorporated into the quoted uncertainties. For simplicity, we refer to any effects associated with these abundance uncertainties as systematic in the following text, although they also include contributions from random noise.

\section{discussion}
\label{discussion}

\subsection{Average variability in each campaign}
Systematic errors make it unclear whether genuine daily variability exists, since a temporally constant abundance could, in principle, show the apparent variability in Fig.~\ref{123ab}. As a first-order test, we examine whether the observed scatter in the daily abundances can be explained by systematic effects alone or whether an additional variance component corresponding to real variability is required. 

To quantify the contribution from systematic effects, we compute, for each of the 102 line-free channels, the standard deviation of the errors across all observing dates within a given campaign. We then adopt the median of the 102 values as an estimate of the systematic scatter. The uncertainty on this estimate is given by the 1-$\sigma$ dispersion of these 102 values, defined by the 16th and 84th percentiles. This systematic scatter is then compared with the standard deviation measured from the observed daily abundances. We estimate the real variability by attributing any excess scatter beyond that expected from systematic effects to a physical origin. Given that the systematic errors are independent of the physical abundances, the real daily variability can be approximated as
\begin{equation}
\sigma_{\mathrm{real}} \approx \sqrt{\sigma_{\mathrm{obs}}^{2} - \sigma_{\mathrm{sys}}^{2}} 
\end{equation}
where $\sigma_{\mathrm{obs}}$ is the standard deviation of the observed daily abundances and $\sigma_{\mathrm{sys}}$ is the systematic scatter estimated from the error maps.

The results are summarised in Table~\ref{var}. For H$_2$O, all three campaigns show tentative evidence of real variability, with the observed scatter marginally exceeding the estimated systematic contribution at a significance of $\lesssim 1\sigma$. SO$_2$ in Campaign~3 shows stronger evidence of real variability, whereas the variability in Campaigns~1 and~2 is more comparable to the systematic level. As these estimates characterise variability across an entire campaign, cases in which $\sigma_{\mathrm{obs}}$ is not significantly larger than $\sigma_{\mathrm{sys}}$ do not necessarily rule out real variability. Instead, this may indicate that the overall variability is moderate, or that significant variations occur only between specific days and are diluted when considering the scatter across the full campaign. For each campaign, the daily variability of both species is estimated over a 1-$\sigma$ range by evaluating $\sigma_{\mathrm{real}}$ at the upper and lower 1-$\sigma$ bounds of $\sigma_{\mathrm{sys}}$.

\begin{table}
\centering
\caption{Decomposition of the observed daily abundance variance into systematic and intrinsic components. Values in brackets indicate the 1-$\sigma$ range.}
\label{var}
\begin{tabular}{lcccc}
\hline
Species & Campaign & $\sigma_{\mathrm{obs}}$ (ppb) & $\sigma_{\mathrm{sys}}$ (ppb) & $\sigma_{\mathrm{real}}$ (ppb) \\
\hline
H$_2$O & 1 & 44 & $41$ ($33$--$51$) & $<29$ \\
H$_2$O & 2 & 62 & $46$ ($36$--$62$) & $<50$ \\
H$_2$O & 3 & 56 & $46$ ($38$--$59$) & $<41$ \\
SO$_2$ & 1 & 6  & $7$ ($5$--$8$)   & $<2$ \\
SO$_2$ & 2 & 7  & $8$ ($6$--$11$) & $<4$ \\
SO$_2$ & 3 & 12 & $9$ ($6$--$12$)  & $4$--$11$ \\
\hline
\end{tabular}
\end{table}

\subsection{Variation between days}
\label{dtd}

\begin{figure*}
\centering

\begin{minipage}[t]{0.31\textwidth}
\centering
\includegraphics[width=\linewidth]{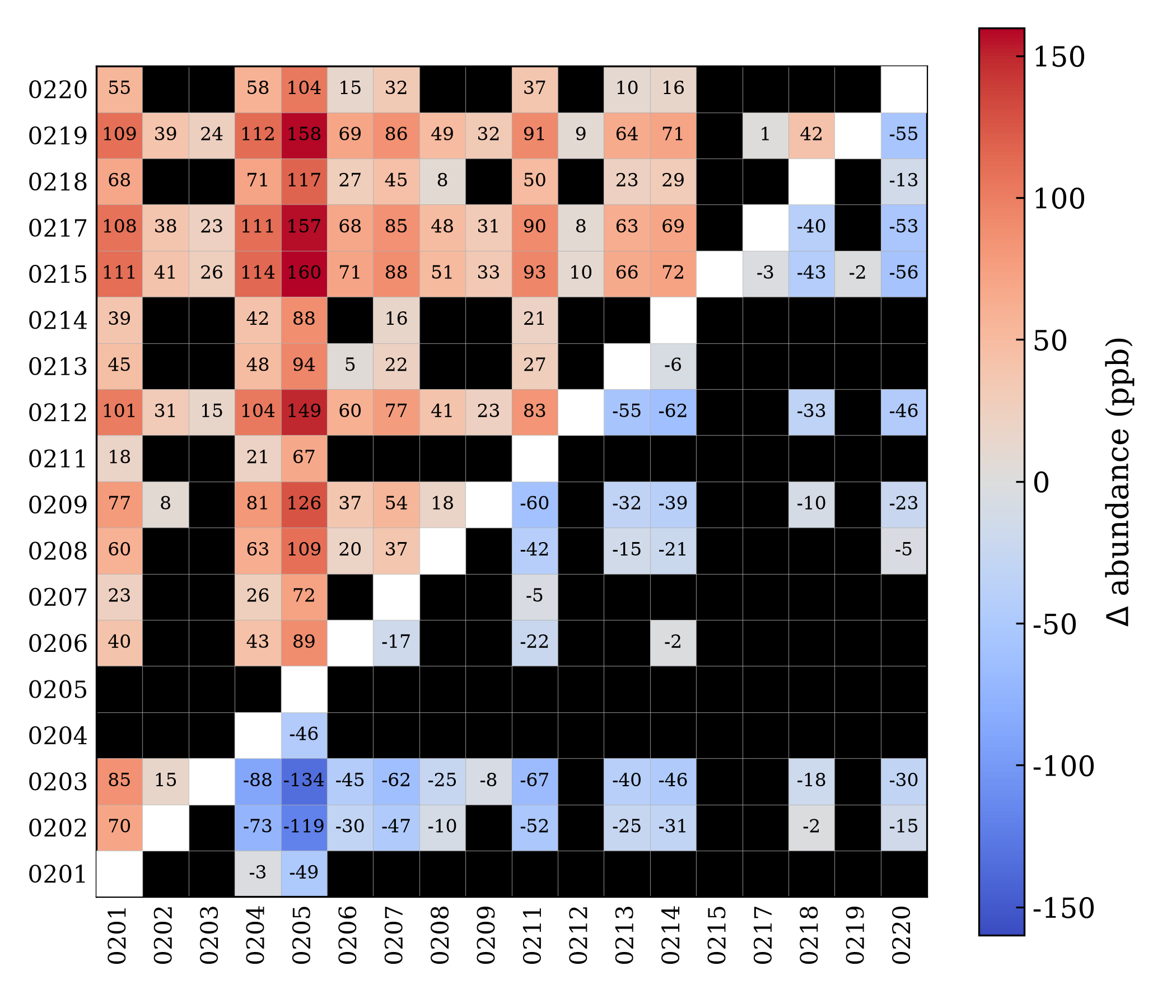}
\par\smallskip
(a) C1: $\Delta$
\end{minipage}\hspace{4pt}
\begin{minipage}[t]{0.31\textwidth}
\centering
\includegraphics[width=\linewidth]{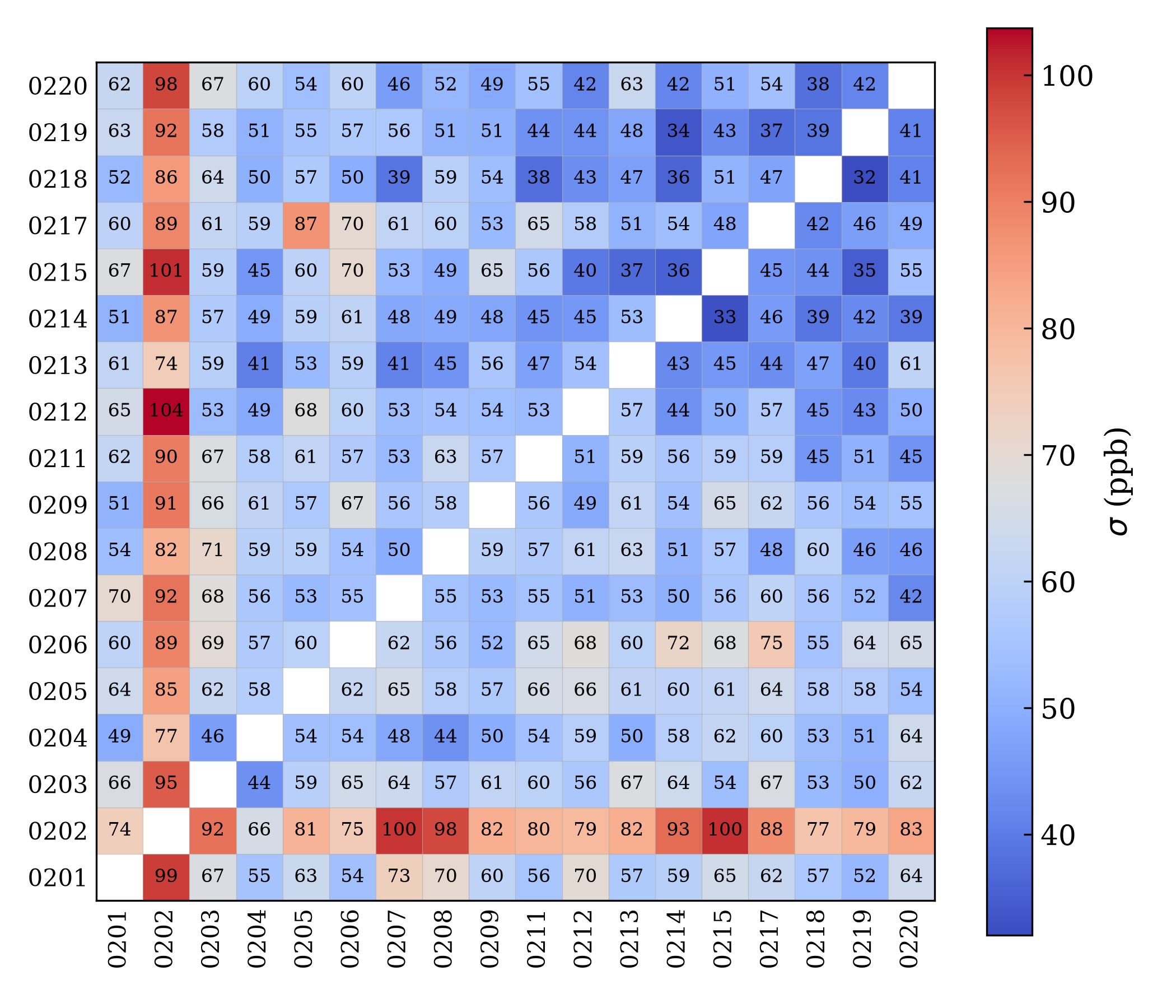}
\par\smallskip
(b) C1: $\sigma$
\end{minipage}\hspace{4pt}
\begin{minipage}[t]{0.31\textwidth}
\centering
\includegraphics[width=\linewidth]{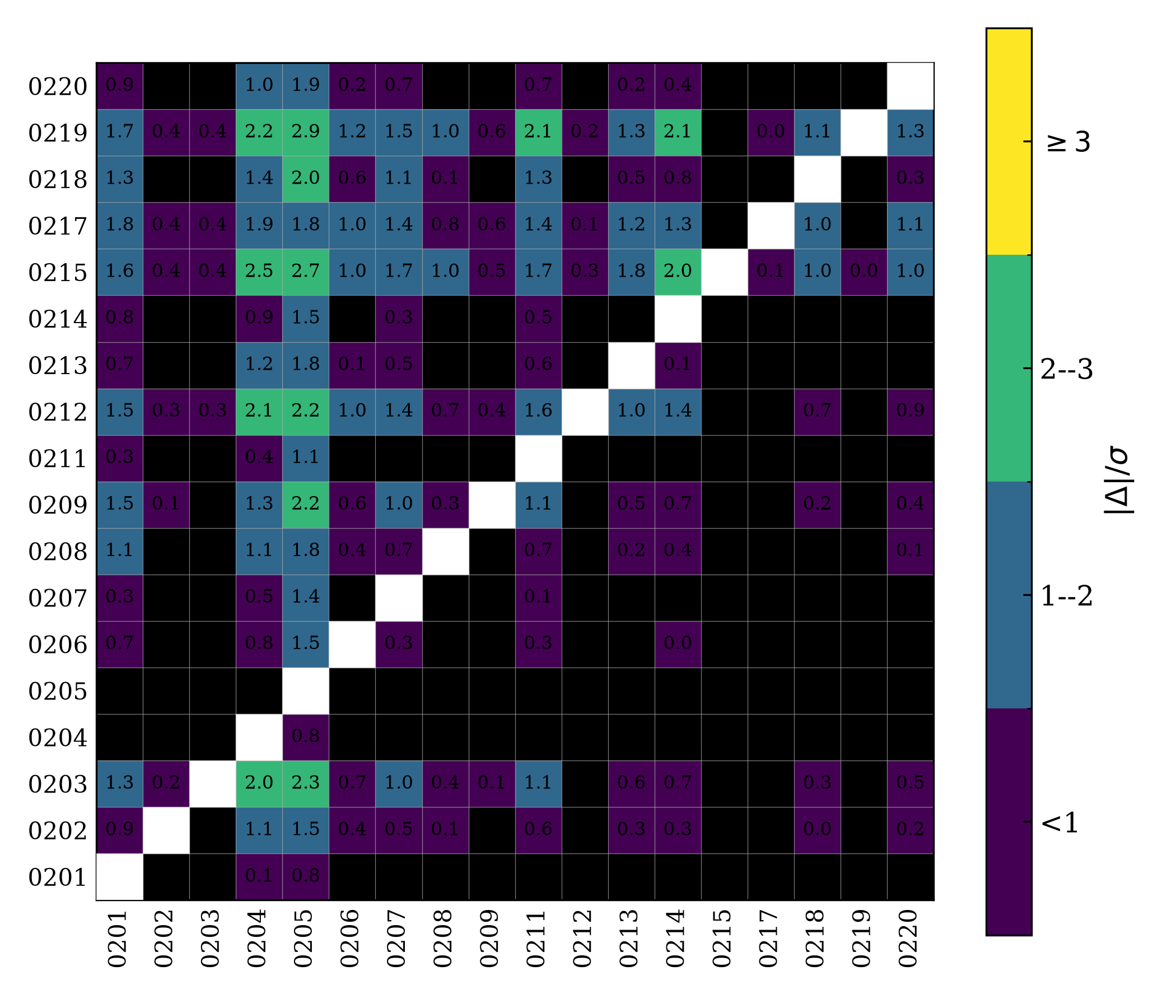}
\par\smallskip
(c) C1: $|\Delta|/\sigma$
\end{minipage}

\vspace{1mm}

\begin{minipage}[t]{0.31\textwidth}
\centering
\includegraphics[width=\linewidth]{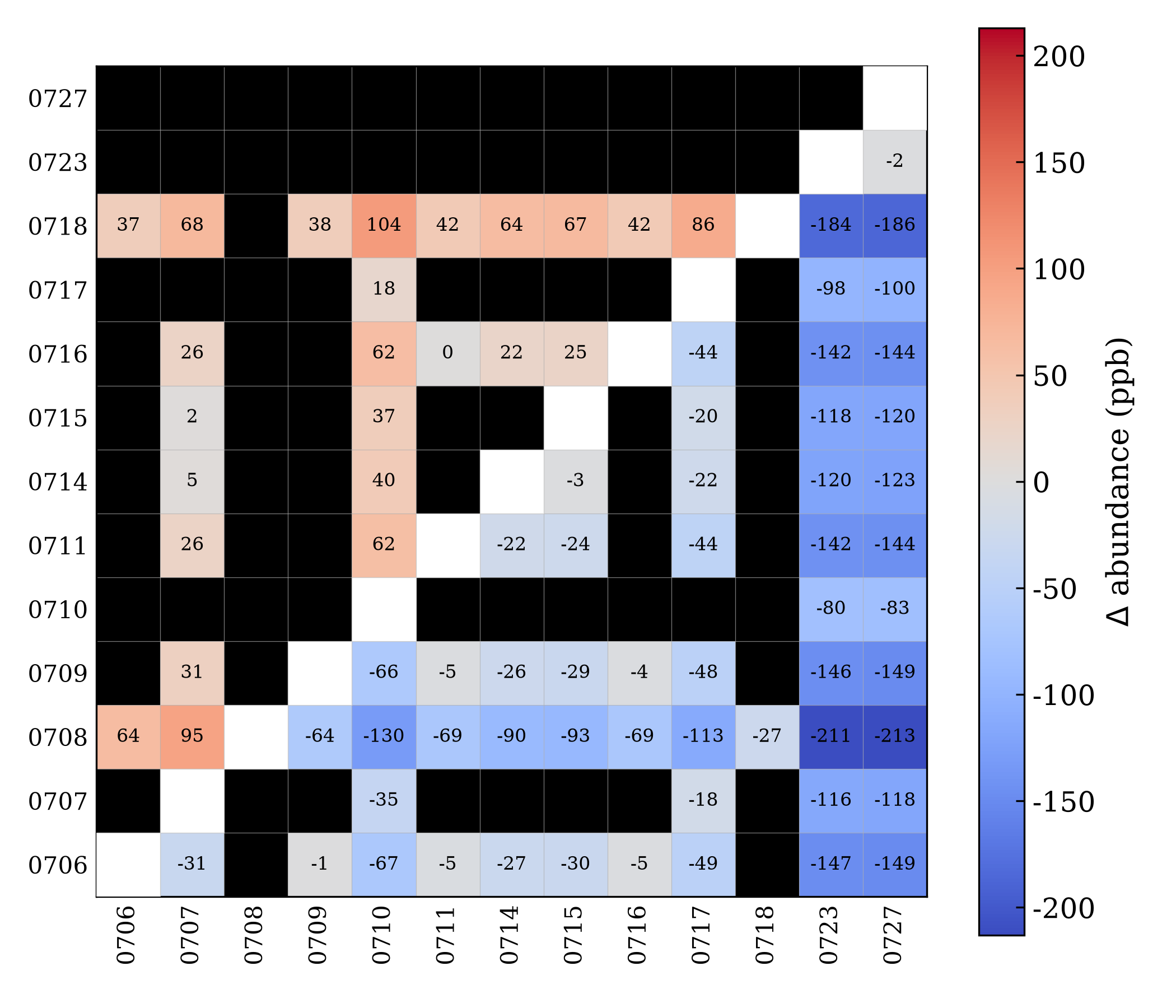}
\par\smallskip
(d) C2: $\Delta$
\end{minipage}\hspace{4pt}
\begin{minipage}[t]{0.31\textwidth}
\centering
\includegraphics[width=\linewidth]{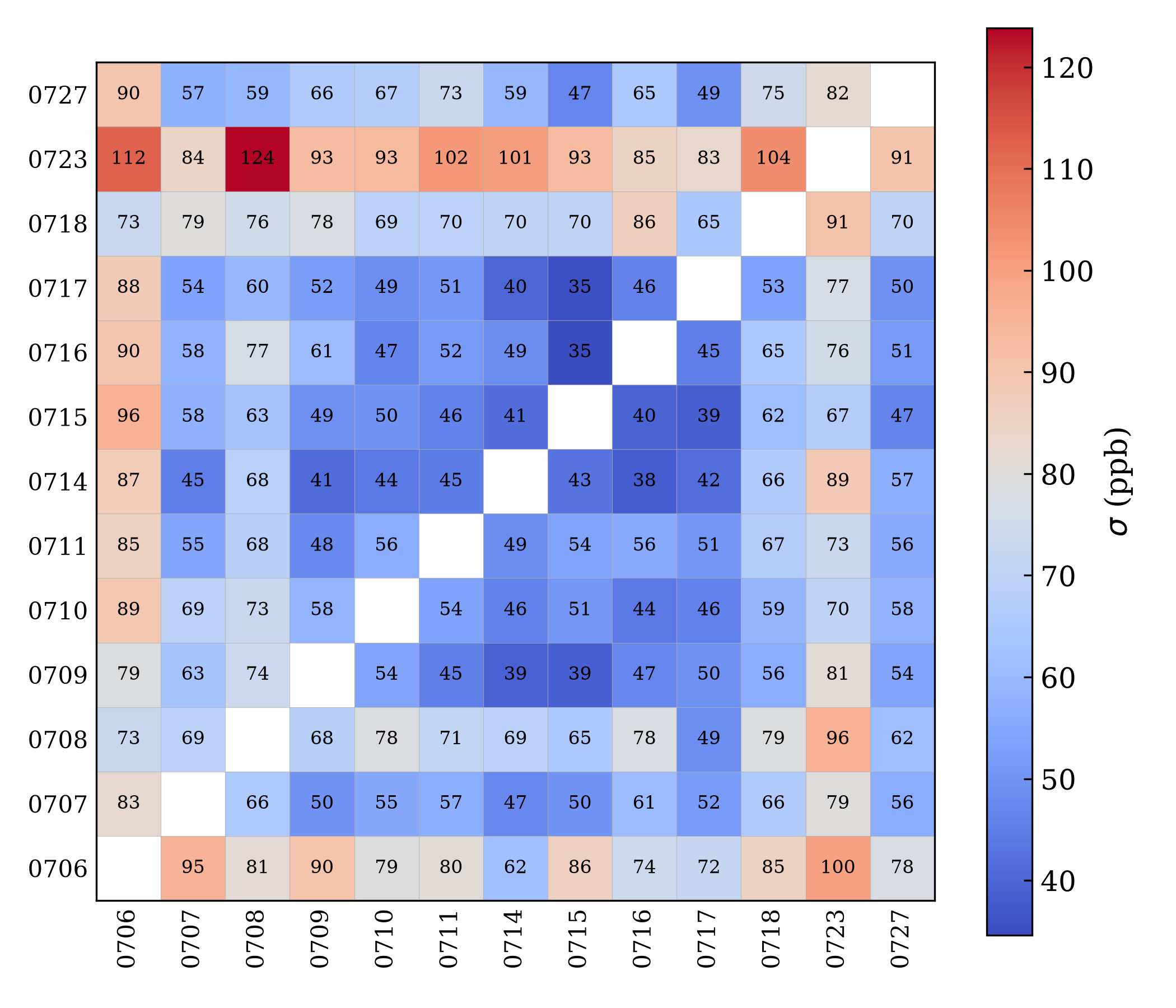}
\par\smallskip
(e) C2: $\sigma$
\end{minipage}\hspace{4pt}
\begin{minipage}[t]{0.31\textwidth}
\centering
\includegraphics[width=\linewidth]{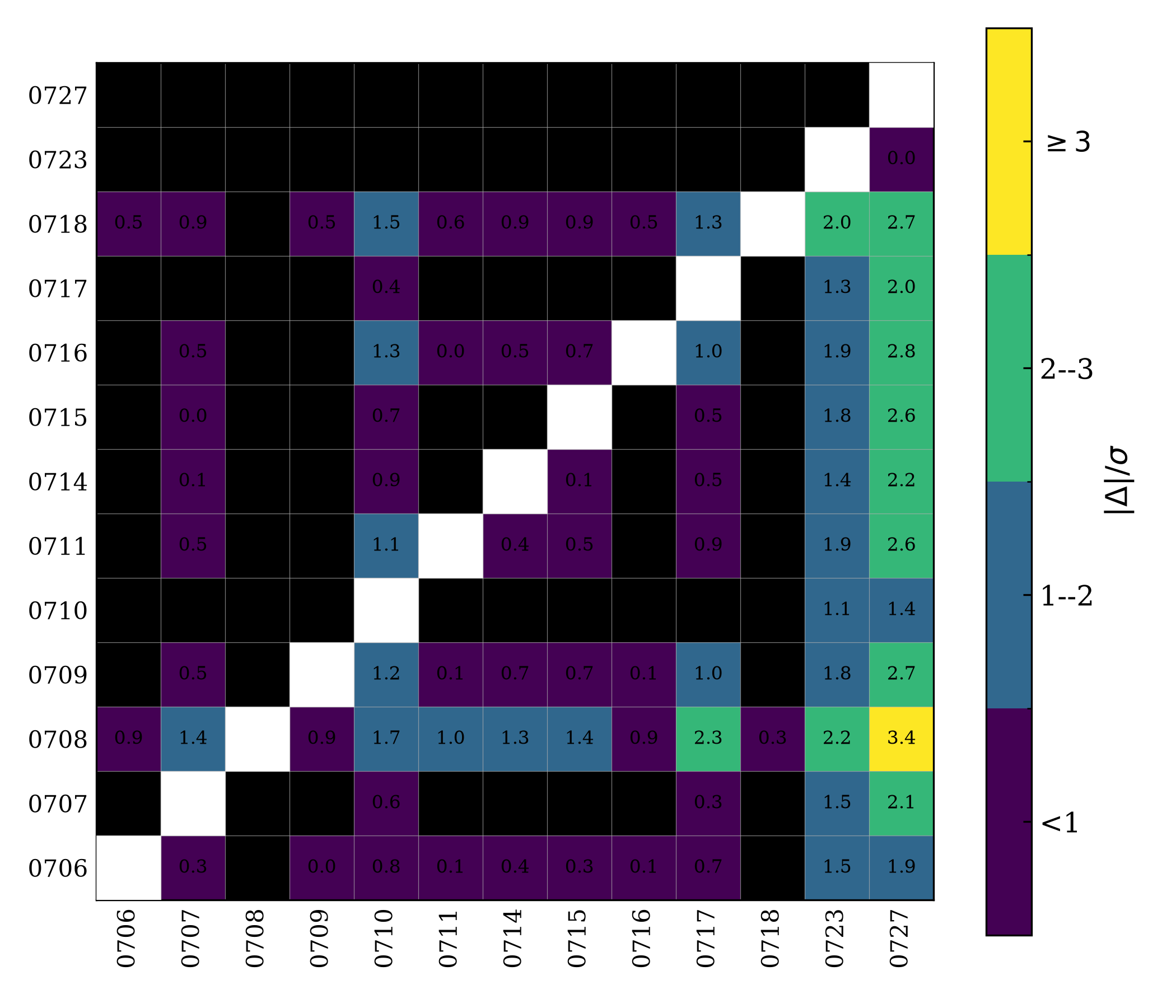}
\par\smallskip
(f) C2: $|\Delta|/\sigma$
\end{minipage}

\vspace{1mm}

\begin{minipage}[t]{0.31\textwidth}
\centering
\includegraphics[width=\linewidth]{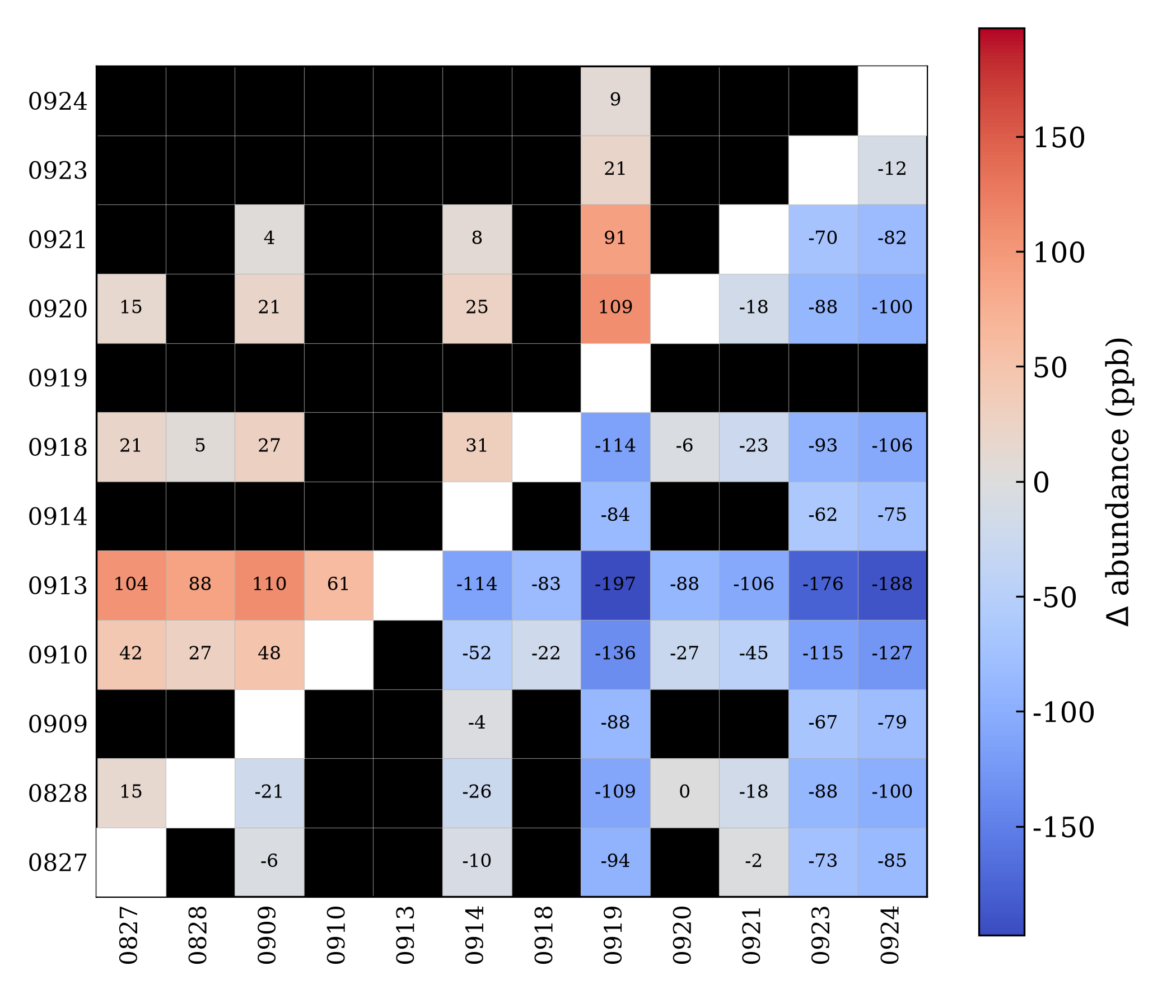}
\par\smallskip
(g) C3: $\Delta$
\end{minipage}\hspace{4pt}
\begin{minipage}[t]{0.31\textwidth}
\centering
\includegraphics[width=\linewidth]{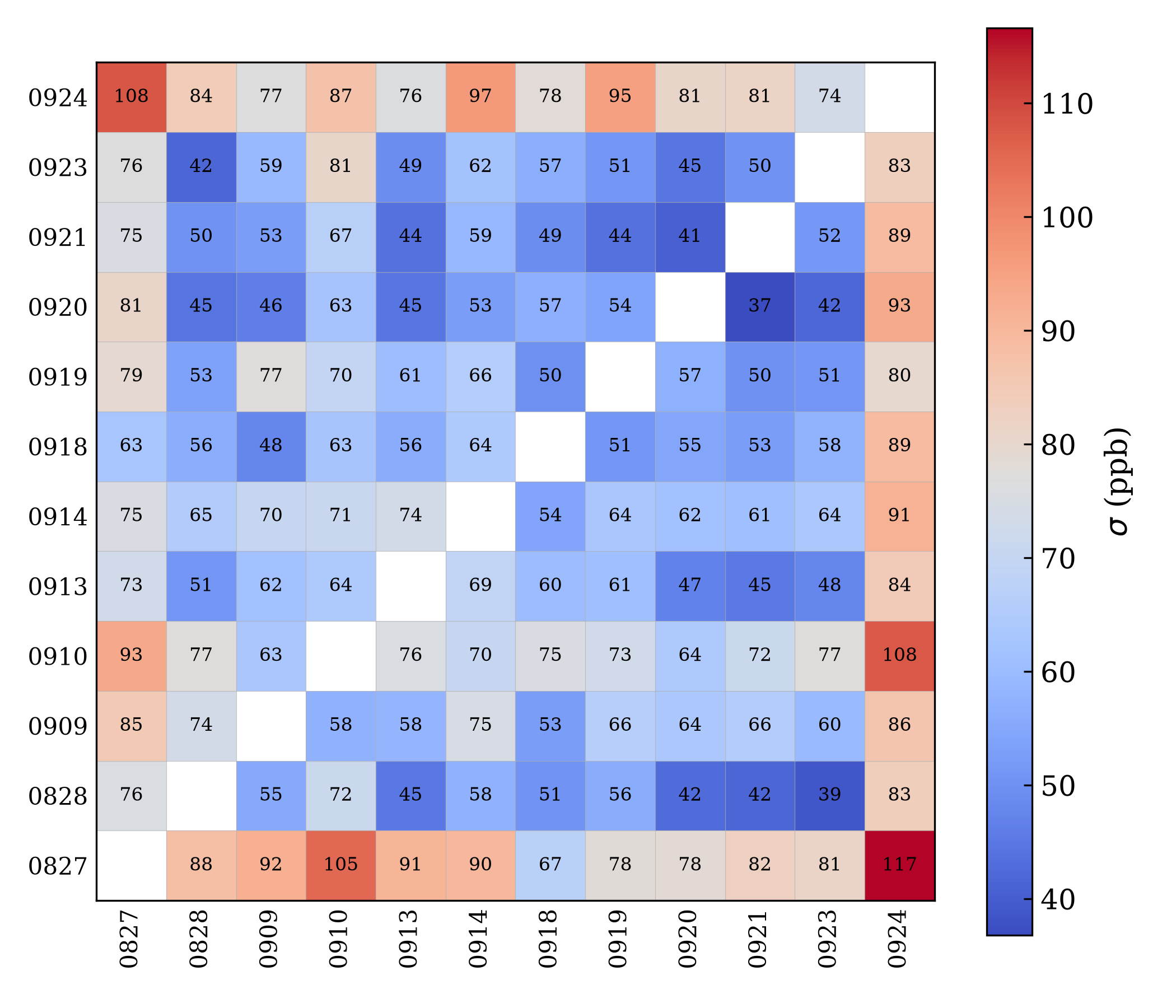}
\par\smallskip
(h) C3: $\sigma$
\end{minipage}\hspace{4pt}
\begin{minipage}[t]{0.31\textwidth}
\centering
\includegraphics[width=\linewidth]{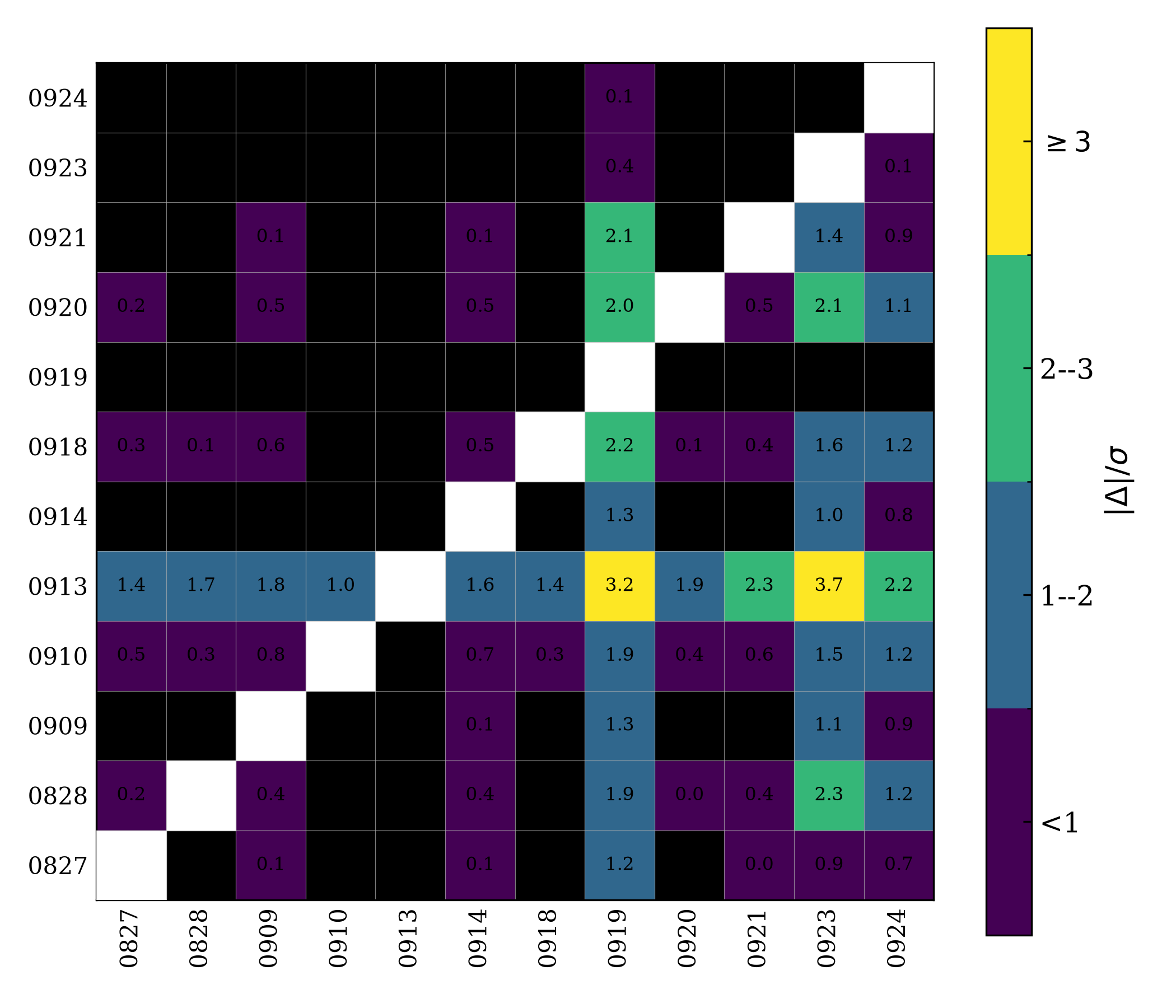}
\par\smallskip
(i) C3: $|\Delta|/\sigma$
\end{minipage}

\caption{
Pairwise H$_2$O matrices for Campaigns~1,~2, and~3. 
Panels show (left) the pairwise abundance-difference matrix $\Delta$,
(middle) the corresponding uncertainty matrix $\sigma$,
and (right) the significance matrix $|\Delta|/\sigma$. In the $\Delta$ panels, red and blue cells indicate abundance increases and decreases, respectively.
}
\label{matrixh2o}
\end{figure*}

\begin{figure*}
\centering

\begin{minipage}[t]{0.31\textwidth}
\centering
\includegraphics[width=\linewidth]{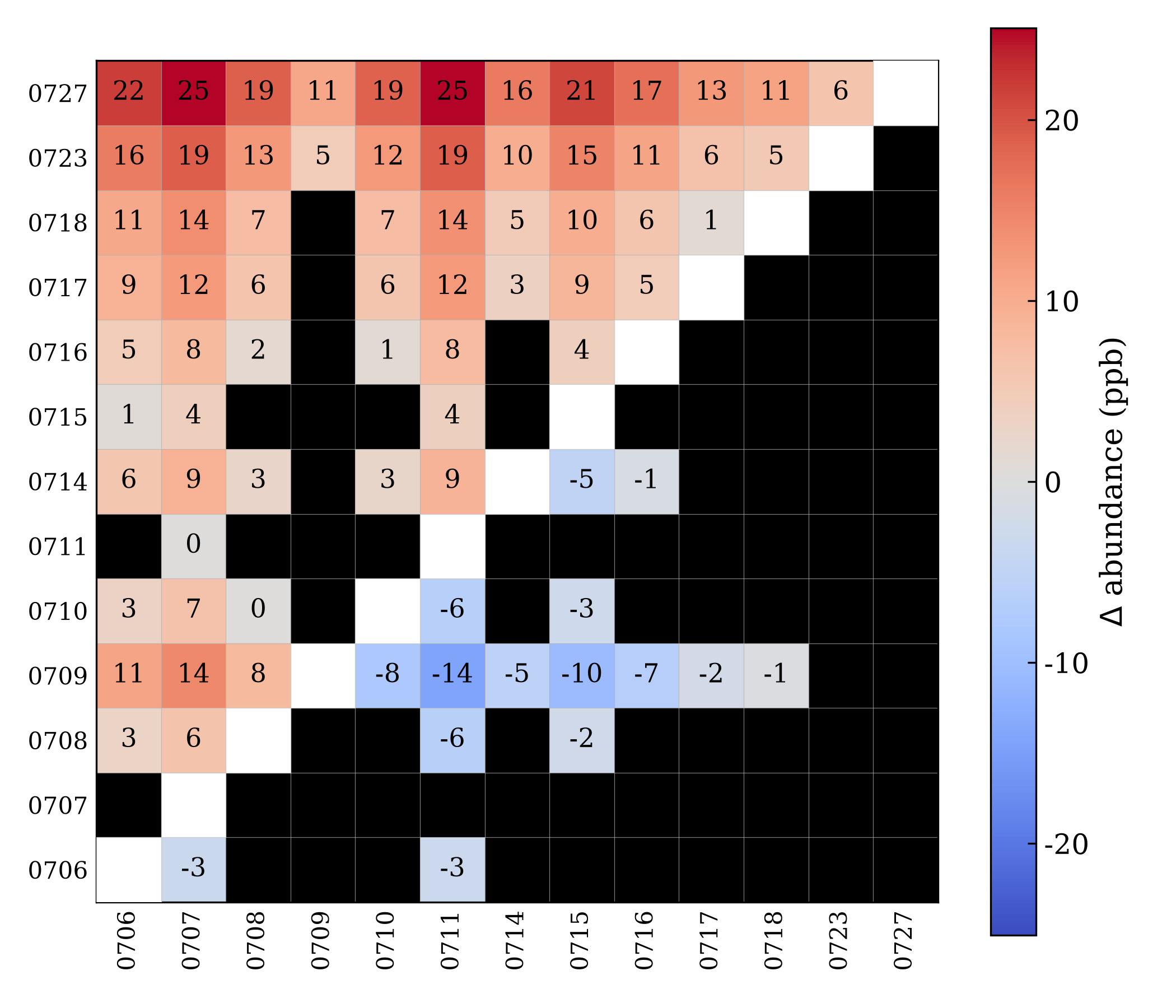}
\par\smallskip
(a) C2: $\Delta$
\end{minipage}\hspace{4pt}
\begin{minipage}[t]{0.31\textwidth}
\centering
\includegraphics[width=\linewidth]{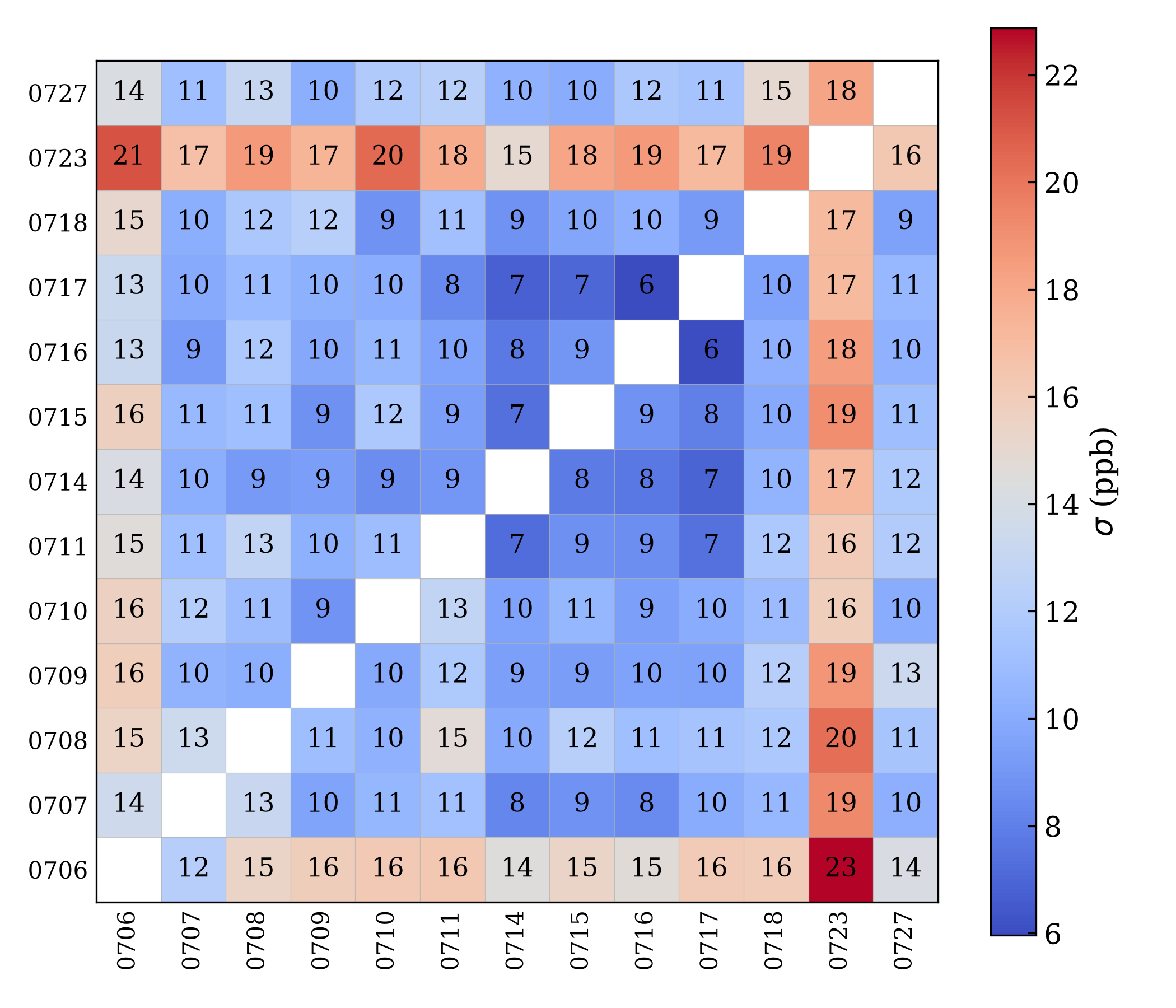}
\par\smallskip
(b) C2: $\sigma$
\end{minipage}\hspace{4pt}
\begin{minipage}[t]{0.31\textwidth}
\centering
\includegraphics[width=\linewidth]{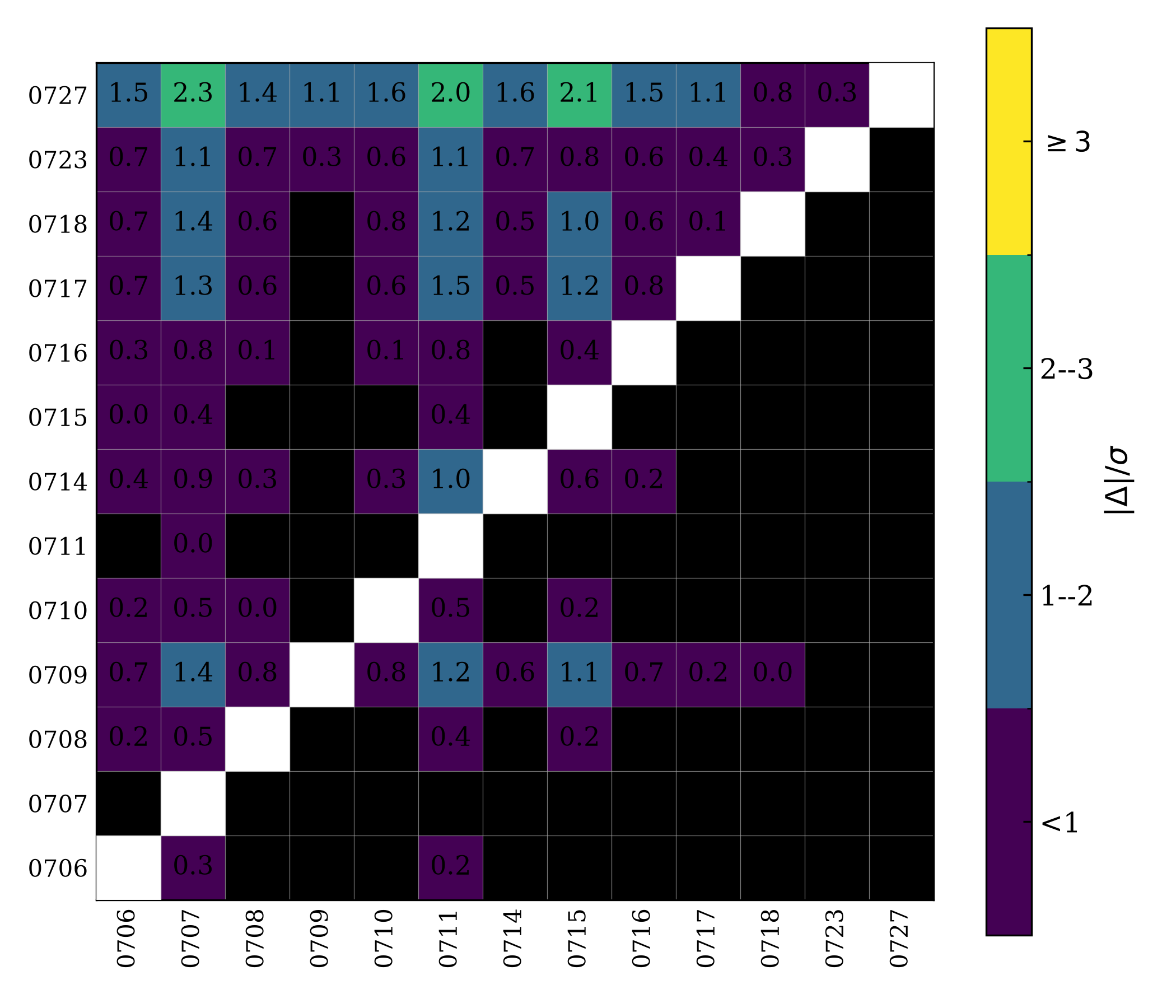}
\par\smallskip
(c) C2: $|\Delta|/\sigma$
\end{minipage}

\vspace{1mm}

\begin{minipage}[t]{0.31\textwidth}
\centering
\includegraphics[width=\linewidth]{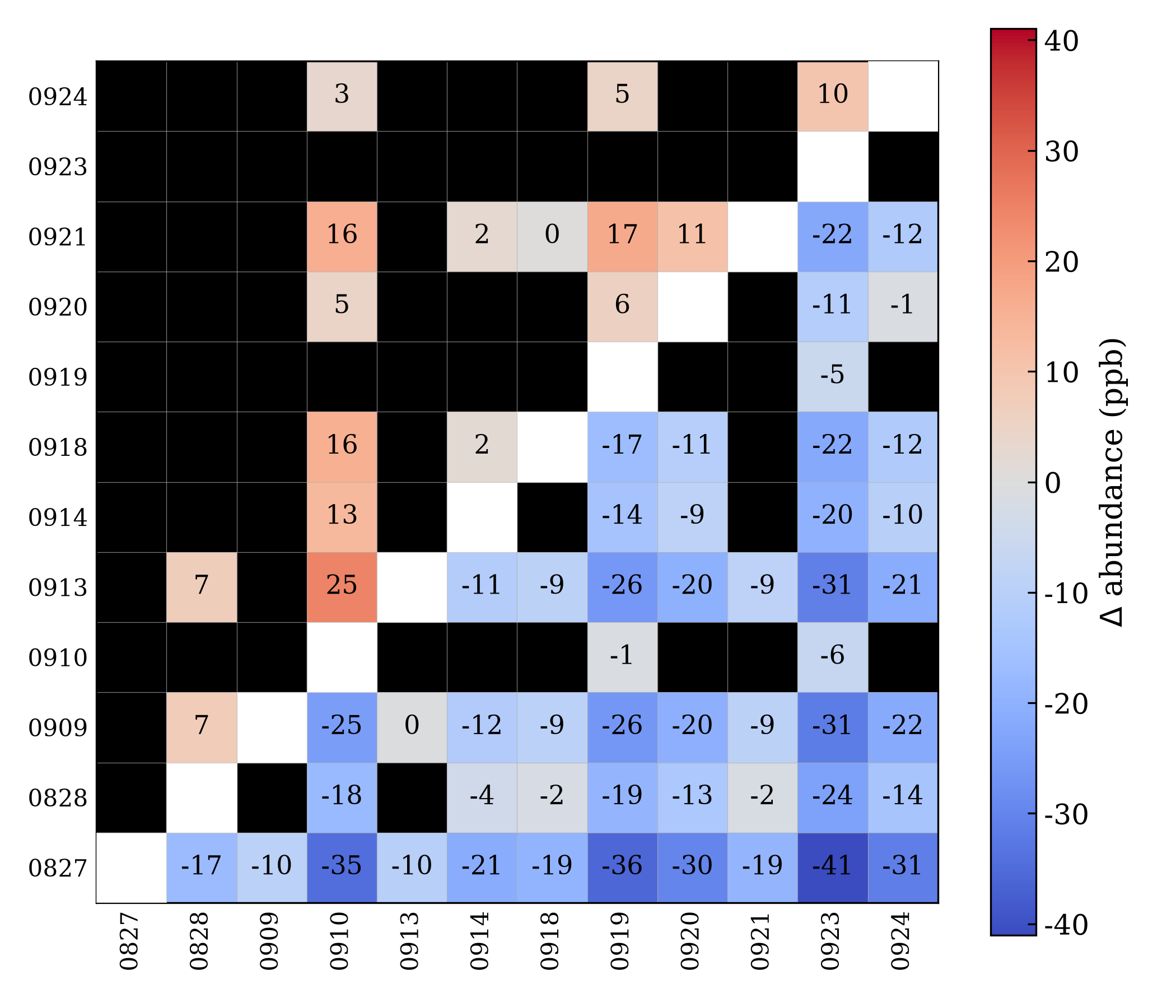}
\par\smallskip
(d) C3: $\Delta$
\end{minipage}\hspace{4pt}
\begin{minipage}[t]{0.31\textwidth}
\centering
\includegraphics[width=\linewidth]{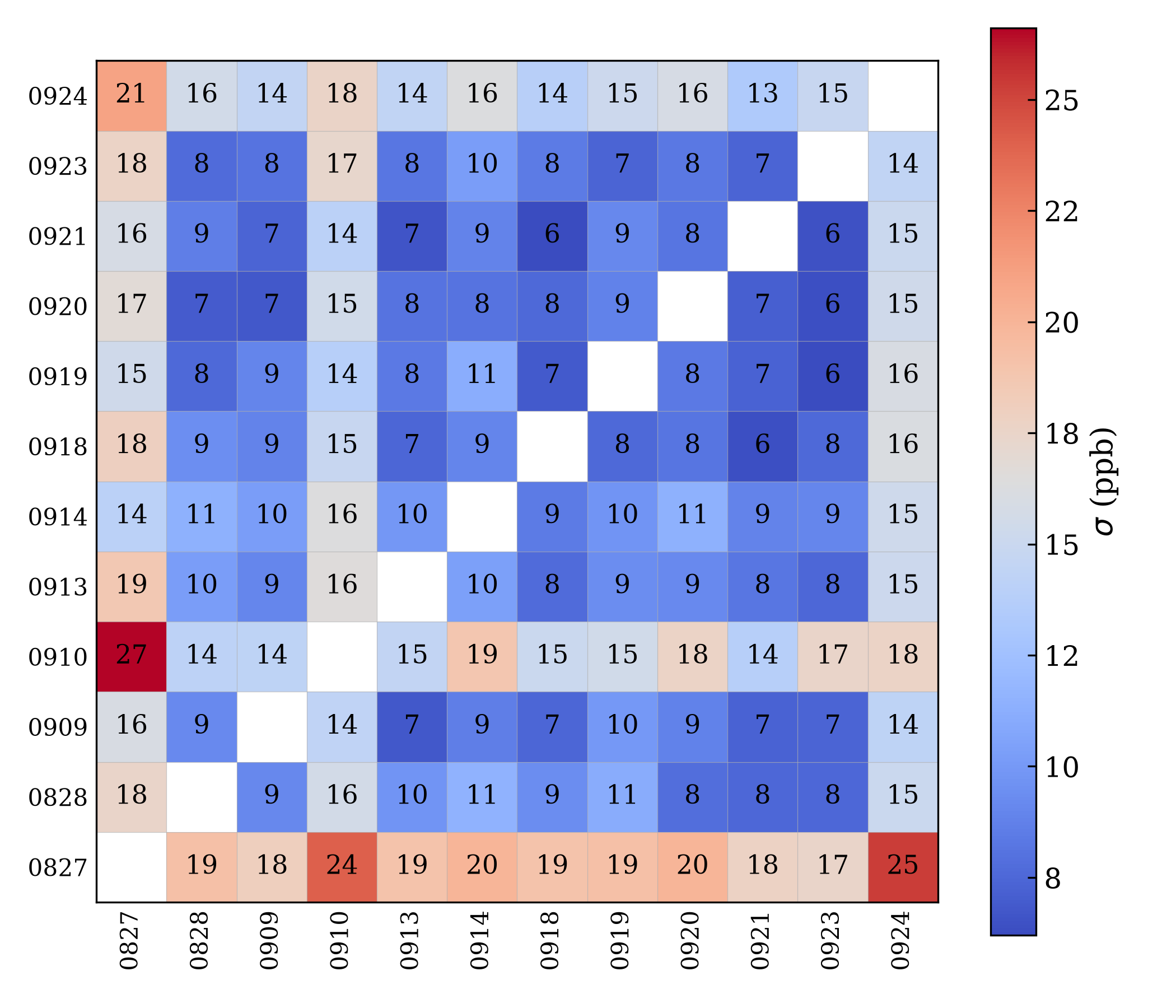}
\par\smallskip
(e) C3: $\sigma$
\end{minipage}\hspace{4pt}
\begin{minipage}[t]{0.31\textwidth}
\centering
\includegraphics[width=\linewidth]{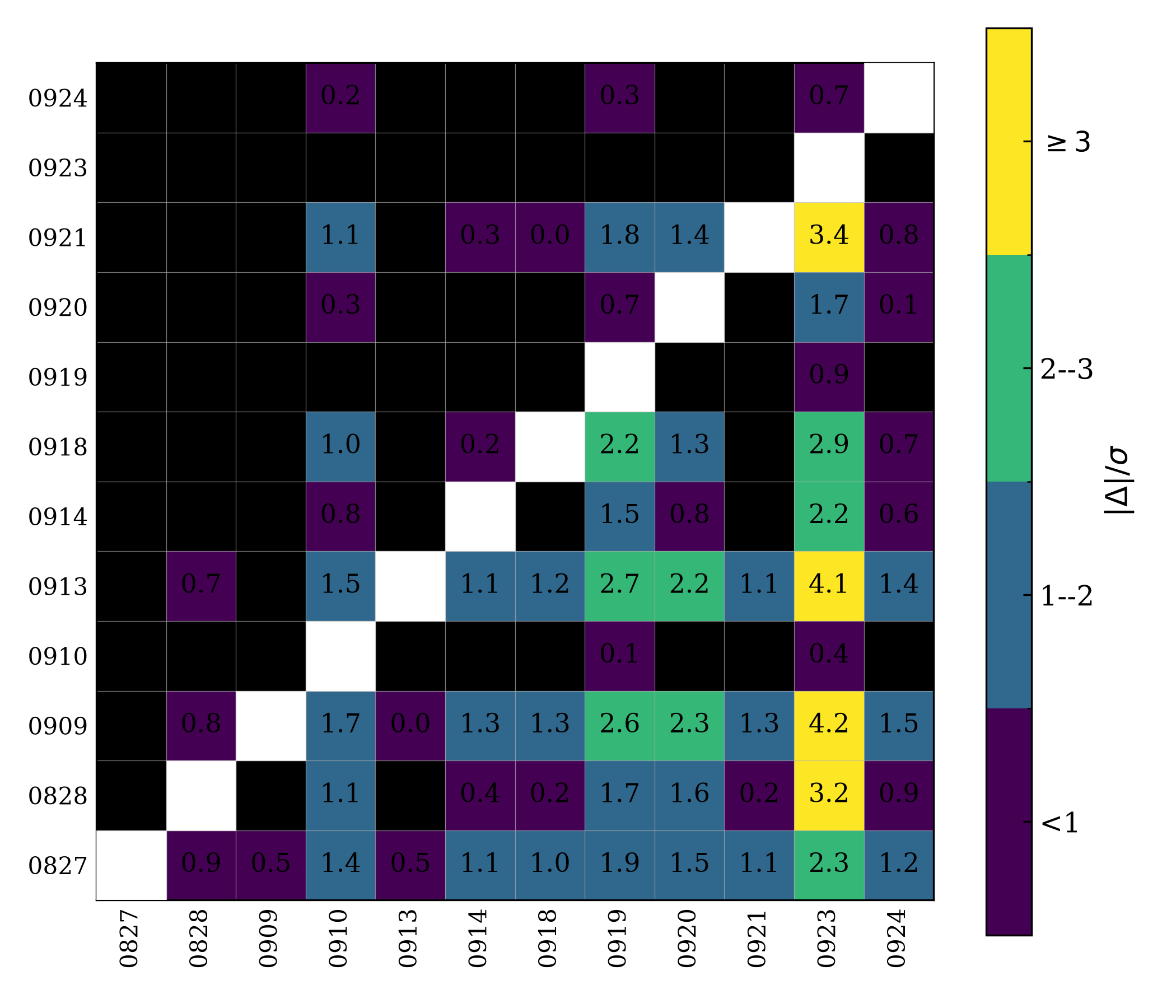}
\par\smallskip
(f) C3: $|\Delta|/\sigma$
\end{minipage}

\caption{
Pairwise SO$_2$ matrices for Campaigns~2 and~3.
Panels show (left) the pairwise abundance-difference matrix $\Delta$,
(middle) the corresponding uncertainty matrix $\sigma$,
and (right) the significance matrix $|\Delta|/\sigma$.
}
\label{matrixso2}
\end{figure*}

After the campaign-scale variability analysis, we further investigate whether variability in excess of the systematic level can be identified on shorter timescales, i.e. between specific observing days within a campaign. For each species and campaign, we construct a variation matrix by pairing all observing dates and computing the abundance difference for each earlier--later date pair. We then quantify the corresponding systematic uncertainty of these variations by constructing an error matrix derived from the abundance error maps. Specifically, for each date pair, we select the corresponding rows of the error map, compute the abundance differences across the 102 line-free channels arising from systematic effects, and determine the 16th and 84th percentiles, which represent the 1-$\sigma$ systematic contributions to the inferred abundance decrease and increase between the two days, respectively. Finally, the abundance variation matrix is normalised by the error matrix to produce a significance matrix, which quantifies the confidence level of the abundance changes between each pair of observing dates. The matrices are shown in Figs.~\ref{matrixh2o} and~\ref{matrixso2}.

For H$_2$O, statistically significant variability at more than 3-$\sigma$ is identified on isolated days in Campaigns~2 and~3. In Campaign~1, no date pair exceeds the 3-$\sigma$ threshold, with only some cells reaching 2--3-$\sigma$. This is consistent with the campaign-averaged analysis, which indicates that the observed scatter exceeds the systematic level only marginally. The 2--3-$\sigma$ cells in Campaign~1 are concentrated in the column of 05 February 2022, suggesting an abundance minimum on that date. Similarly, in Campaign~2, a significant decline is detected on 27 July 2023, when no line detection is obtained, with the corresponding column dominated by 2--3-$\sigma$ values. Although the line is also undetected on 23 July, the larger uncertainties on that date reduce the significance of the associated variations. In Campaign~3, two significant cells are associated with 13 September, while the corresponding row is also dominated by high-significance values, indicating an abundance peak on that date.

Overall, the slightly enhanced observed variability of H$_2$O relative to the systematic level in all three campaigns is unlikely to be driven by a smooth campaign-wide evolution, and instead appears to arise from a generally stable background punctuated by temporary fluctuations. Specifically, the apparent minimum on 05 February 2022, the depletion on 27 July 2023, and the enhancement on 13 September 2023 appear to account for much of the scatter in excess of the systematic level within their respective campaigns. This interpretation is supported by recomputing the campaign-averaged variability after removing those days showing abrupt excursions (Table~\ref{var2}). The estimated systematic contribution changes negligibly, whereas the observed scatter, and hence the inferred real variability, decreases noticeably. The slight excess variability remaining after removal of these excursions in all three campaigns may indicate the presence of additional small-amplitude fluctuations, which are constrained to an average level below $\sim$40~ppb. This is consistent with the results of \citet{bertaux2007warm} and \citet{fedorova2008hdo}, who found little variability, while also remaining compatible with studies that reported strong variability associated with sudden depletions, for which sparser sampling may simply have captured days such as 27 July \citep{sandor2005water,gurwell2007swas,chamberlain2020soir}.

For SO$_2$, no statistically significant variations are detected in Campaign~2 (Fig.~\ref{matrixso2}). However, the variation matrix preferentially populates the increase region and exhibits a triangular structure, with the magnitude of the variation generally increasing with temporal separation. This pattern is consistent with a gradual monotonic increase in abundance across the campaign, whose amplitude remains insufficient for any individual date pair to exceed the 3-$\sigma$ threshold. By contrast, multiple date pairs in Campaign~3 exceed the 3-$\sigma$ threshold, and all correspond to decreases in abundance. Cells above the 2-$\sigma$ level are likewise concentrated predominantly in the decrease region of the matrix. Moreover, the high-significance cells are primarily associated with later observing dates, in other words clustering towards the right-hand side of the matrix, rather than forming a smoothly distributed triangular pattern. The pattern indicates a relatively stable initial phase followed by a later reduction, rather than a monotonic decline across the entire campaign. Short-term variability also appears to be present, most notably in the variation between 21 and 23 September. A more quantitative assessment of the campaign-wide trend is presented in Appendix~\ref{mt}.

\begin{table}
\centering
\caption{
Same as Table~\ref{var}. Bold values indicate the results obtained before excluding the specified observing dates.
}
\label{var2}
\begin{tabular}{lcccc}
\hline
Species & Campaign & $\sigma_{\mathrm{obs}}$ (ppb) & $\sigma_{\mathrm{sys}}$ (ppb) & $\sigma_{\mathrm{real}}$ (ppb) \\
\hline
H$_2$O & 1 & 37 & $41$ ($32$--$51$) & $<19$ \\
      &   & $\mathbf{44}$ & $\mathbf{41}$ ($\mathbf{33}$--$\mathbf{51}$) & $<\mathbf{29}$ \\[2pt]
H$_2$O & 2 & 53 & $46$ ($35$--$64$) & $<40$ \\
      &   & $\mathbf{62}$ & $\mathbf{46}$ ($\mathbf{36}$--$\mathbf{62}$) & $<\mathbf{50}$ \\[2pt]
H$_2$O & 3 & 46 & $47$ ($37$--$61$) & $<28$ \\
      &   & $\mathbf{56}$ & $\mathbf{46}$ ($\mathbf{38}$--$\mathbf{59}$) & $<\mathbf{41}$ \\
\hline
\end{tabular}
\end{table}

\subsection{Variation between campaigns}
\label{ctc}

We determine the mean abundances and associated uncertainties for each campaign, and present these in Table~\ref{mean}. These values are derived using the same procedure as for the daily measurements, except for the additional step of averaging all daily spectra within each campaign. Campaign~1 SO$_2$ is a non-detection overall.

Following the same logic described in Section~\ref{dtd}, we determine the significance of the variation between the three campaigns. For H$_2$O, the mean abundance decreases from 154~ppb in Campaign~1 to 65 and 89~ppb in Campaigns~2 and 3, respectively. These decreases correspond to significances of 3.4-$\sigma$ and 3.0-$\sigma$, respectively, indicating real variation, whereas the difference between Campaigns~2 and 3 is only 1.0-$\sigma$ and is therefore consistent with systematic effects. For SO$_2$, the mean abundance increase between Campaigns~2 and 3 is significant at the 5.7-$\sigma$ level, likewise indicating real variation. The differences between Campaign~1 and Campaigns~2 and 3 are significant at the 4.9-$\sigma$ and 12.9-$\sigma$ levels, respectively, further implying that SO$_2$ is genuinely absent in Campaign~1.

This suggests that H$_2$O is broadly stable between Campaign~2 in July 2023 and Campaign~3 in late August to September 2023, corresponding to a timescale of roughly seven weeks. Meanwhile, a long-term decrease is evident from Campaign~1 in February 2022 to the two campaigns in 2023. Previous studies have not clearly identified such long-term variability. This is mainly attributed to observational limitations: sub-mm/mm observations were too sparsely sampled to separate longer-term behaviour from short-term variability \citep{sandor2005water,gurwell2007swas}, while solar-occultation measurements were limited by irregular sampling and changing latitudinal coverage \citep{chamberlain2020soir}. Our observations are less affected by these limitations because they are disc-integrated, and each campaign spans several weeks with a relatively dense sampling rate. This allows the campaign-averaged abundances to be regarded as representative of each observing epoch, reducing the likelihood that the differences between epochs are driven by anomalously high or low abundances on individual days. This interpretation is further strengthened by the finding that the H$_2$O abundances are relatively stable within each campaign.

Analogous to the argument for H$_2$O, the absence of SO$_2$ in Campaign~1 and its appearance in Campaigns~2 and 3 likewise indicate clear long-term variation. Such long-term variability in the upper mesosphere has previously been suggested only by \citet{belyaev2017night}, and no clear trend, like the decadal variation observed at the cloud tops, has yet been established. \citet{encrenaz2025hdo}, whose observations are sensitive to the cloud tops, obtain measurements in February 2022, July 2023, and October 2023 that are nearly simultaneous to our three campaigns. They likewise find lower SO$_2$ abundances in February 2022, of about 100~ppb, and higher abundances in July and October 2023, of about 450~ppb. This suggests that the SO$_2$ abundance in the upper mesosphere may be directly related to the abundance at the cloud tops, and may follow the same long-term trend. In other words, the long-term behaviour in the upper mesosphere may be modulated by the supply of SO$_2$ from the cloud tops.

At the cloud tops, the H$_2$O abundance is found to be comparable across the three campaign periods \citep{encrenaz2025hdo}, so no significant long-term difference would be expected from changes in upward transport alone. This in turn suggests that the long-term variation of H$_2$O seen in the upper mesosphere is more likely to represent a response to the change in SO$_2$, through H$_2$SO$_4$ formation. A lower cloud-top SO$_2$ abundance, as in Campaign~1, may result in an insufficient upward flux to maintain detectable SO$_2$ in the upper mesosphere, thereby allowing a higher H$_2$O abundance to persist there. As the cloud-top SO$_2$ abundance increases along the long-term trend, as in Campaigns~2 and 3, the upward supply to the upper mesosphere may also increase, raising the SO$_2$ abundance there while consuming more H$_2$O through H$_2$SO$_4$ formation. {This upper-mesospheric anticorrelation, modulated by supply from lower altitudes, has been predicted by the sulphur--water chemistry model of \citet{shao2020revisiting}.}

One of the most notable anomalies in the upper mesosphere is the SO$_2$ vertical inversion, in which the SO$_2$ abundance decreases with altitude to depletion by $\sim$80--85 km and then increases again above this level \citep{sandor2010sulfur,belyaev2012vertical}. This suggests the presence of an additional SO$_2$ reservoir in the upper mesosphere, whose origin remains under debate \citep{vandaele2017sulfur1}. Proposed candidates include the evaporation of H$_2$SO$_4$ aerosol followed by photodissociation \citep{zhang2010photolysis} and the oxidation of polysulphur (S$_x$) produced by OCS photodissociation near the cloud tops and subsequently transported upward \citep{zhang2012sulfur}. Alternatively, the inversion may arise purely from dynamics, namely a mesospheric circulation analogous to the Hadley cell, which could produce an inversion at middle and high latitudes \citep{stolzenbach2015three}.

If our proposed link between the long-term behaviour of SO$_2$ in the upper mesosphere and at the cloud tops is correct, this would imply that the unknown reservoir is likewise tied to the supply of SO$_2$ or a chemically related species from lower altitudes. Otherwise, detectable SO$_2$ would be expected to persist irrespective of changes below. This would make polysulphur derived from OCS a less likely explanation, unless OCS also exhibits a long-term trend positively correlated with SO$_2$. Photodissociation of H$_2$SO$_4$ vapour remains a possibility, especially in light of the earlier Pioneer Venus result that H$_2$SO$_4$ haze and SO$_2$ show a strong positive correlation in their long-term behaviour \citep{esposito1988sulfur}. However, this model requires substantially more H$_2$SO$_4$ vapour than has been observationally established, for which only upper limits are available \citep{sandor2012upper}, and also relies on extreme assumptions for other parameters \citep{zhang2010photolysis}. We will further test this scenario in \citet{Tang2026}, which investigates the sulphur budget in the upper mesosphere, including H$_2$SO$_4$, together with SO and OCS, over our three campaigns. Finally, the dynamical model of \citet{stolzenbach2015three} is not in conflict with our finding and therefore remains a possible explanation.

\begin{table}
\centering
\caption{Mean abundances of H$_2$O and SO$_2$ in the three campaigns.}
\label{mean}
\begin{tabular}{lccc}
\hline
Species & Campaign & Abundance (ppb) & Error (ppb) \\
\hline
H$_2$O & 1 & 154 & $+15/-17$ \\
H$_2$O & 2 & 65 & $+15/-17$ \\
H$_2$O & 3 & 89 & $+14/-15$ \\
SO$_2$ & 1 & 0 & $+2/-2$ \\
SO$_2$ & 2 & 23 & $+3/-3$ \\
SO$_2$ & 3 & 46 & $+3/-3$ \\
\hline
\end{tabular}
\end{table}

\subsection{Diurnal cycle}

Considering our observing geometry for Campaigns~2 and 3, both span LST ranges with one edge fixed near midnight, while the opposite edge extends progressively into the dayside. The key difference lies in the direction of this extension: in Campaign~2, the coverage extends from the evening terminator into the dayside, whereas in Campaign~3 it extends from the morning terminator into the dayside. This complementary geometry provides a favourable configuration for testing LST dependence and even diurnal asymmetries in both species. The analysis of diurnal variation based on Campaigns~2 and 3 is made feasible by the absence of any long-term difference over this period, which is supported by the broadly stable supplies of SO$_2$ and H$_2$O from lower altitudes inferred from the near-simultaneous observations of \citet{encrenaz2025hdo}, and further supported by the nearly consistent H$_2$O abundance between the two campaigns as discussed in Section~\ref{ctc}. In contrast, the use of Campaign~1 for LST analysis is hindered by long-term behaviour, although it samples a geometry similar to that of Campaign~3.

\begin{figure*}
    \centering

    \includegraphics[width=.9\textwidth]{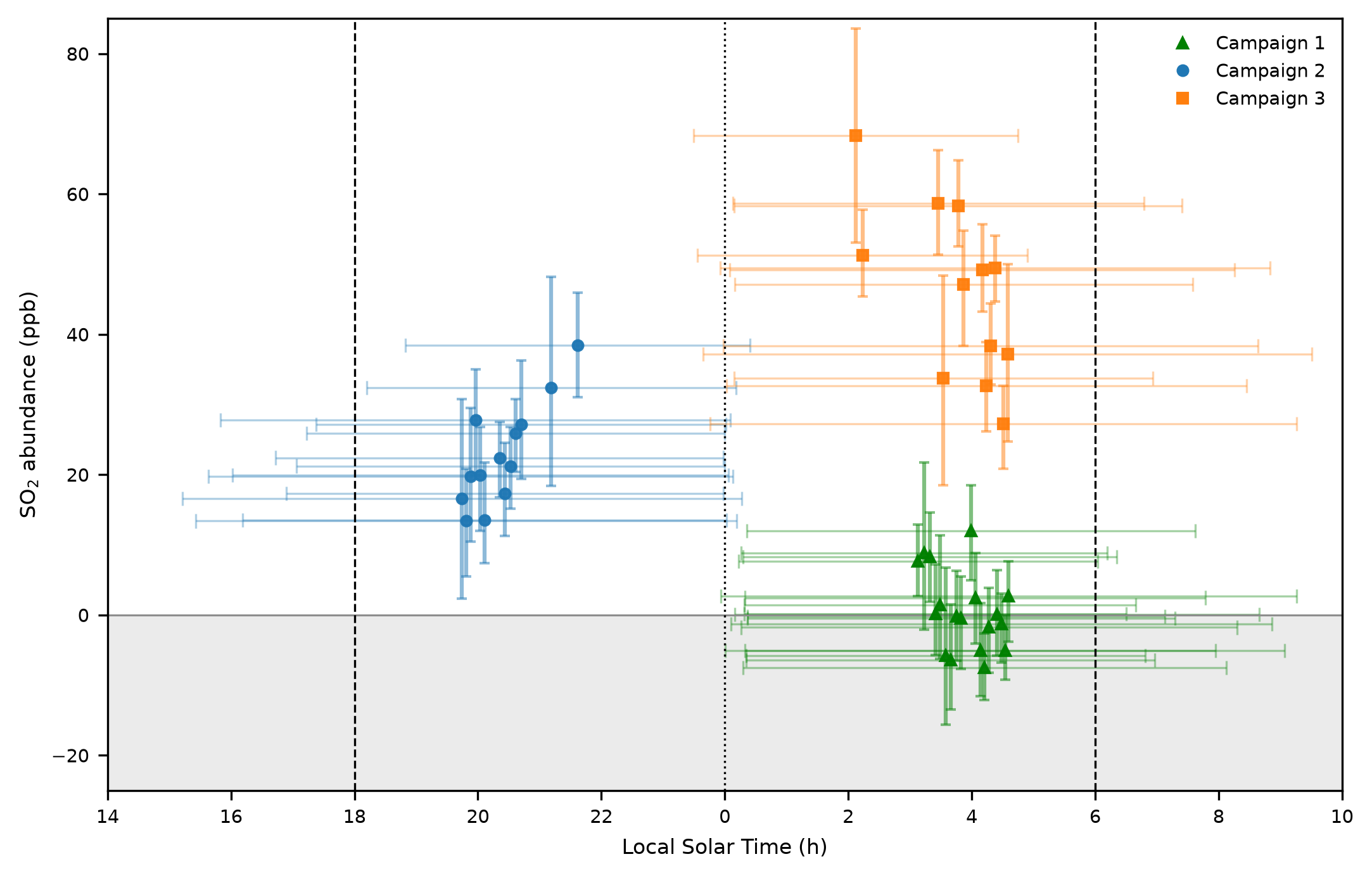}

    \vspace{0.3cm}

    \includegraphics[width=.9\textwidth]{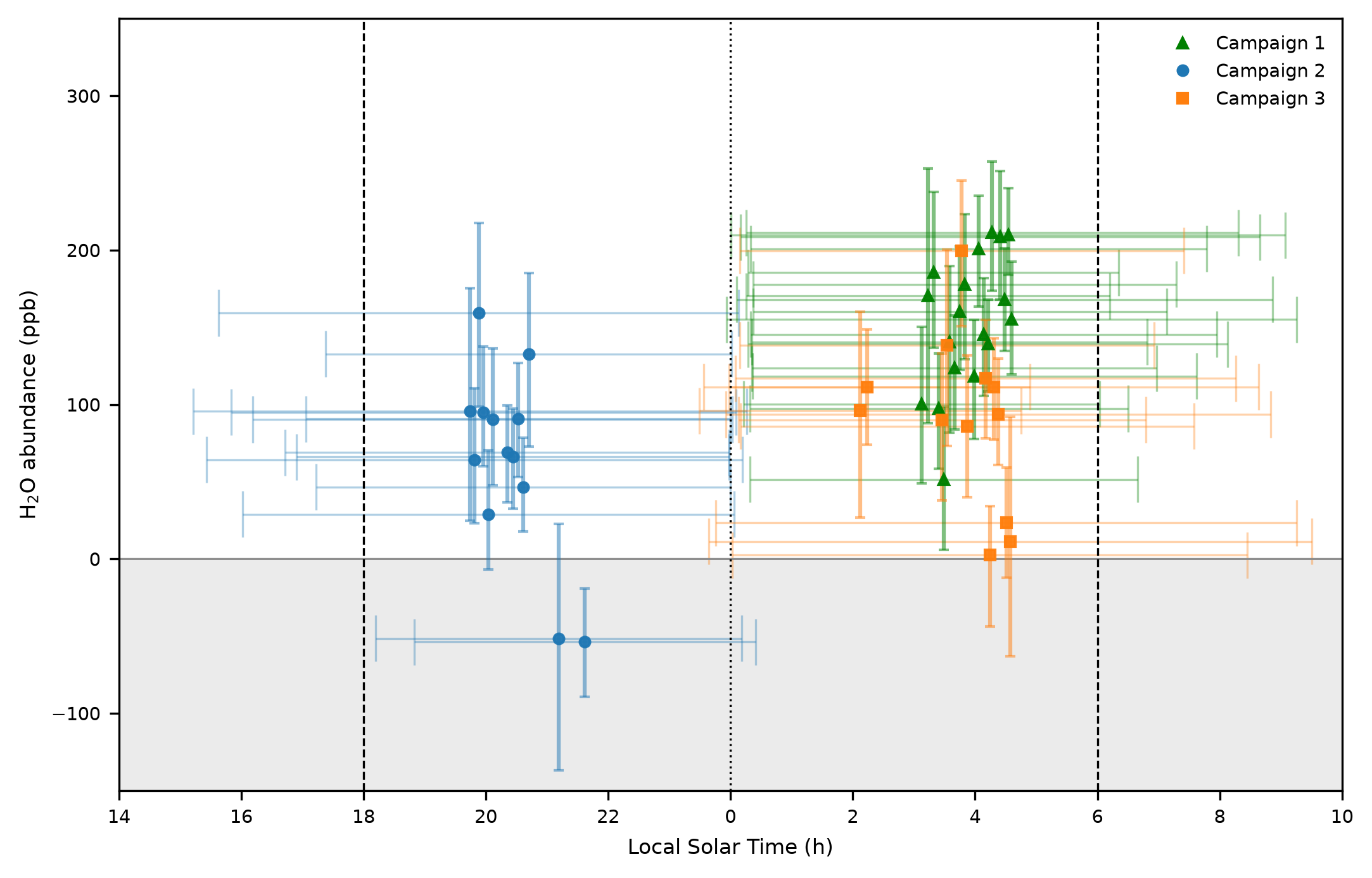}

   \caption{
Observed disc-integrated abundances as a function of the sampled LST range. 
Top: SO$_2$. Bottom: H$_2$O. Horizontal bars indicate the LST coverage of each observing day corresponding to the 90\% beam power. 
The markers represent the disc-integrated abundances and are placed at the sub-observer LST, corresponding to the peak beam response, and the vertical error bars denote the 1-$\sigma$ uncertainties. 
Green triangles indicate Campaign~1, blue circles Campaign~2, and orange squares Campaign~3. }

    \label{so2_lst}
\end{figure*}

As discussed in Sections~\ref{dtd} and \ref{ctc}, we detect a difference in the campaign-averaged SO$_2$ abundance between Campaigns~2 and 3, and identify campaign-wide trends within each, characterised by a monotonic increase during Campaign~2 and by initially stable abundances followed by a decline later in Campaign~3. We interpret these features as being linked to the evolution of the sampled LST, and illustrate them by plotting the abundance as a function of LST coverage in Fig.~\ref{so2_lst}. The Campaign~1 data are also shown, clearly demonstrating the long-term variation in both H$_2$O and SO$_2$.

\begin{figure*}
\centering
\includegraphics[width=0.92\textwidth]{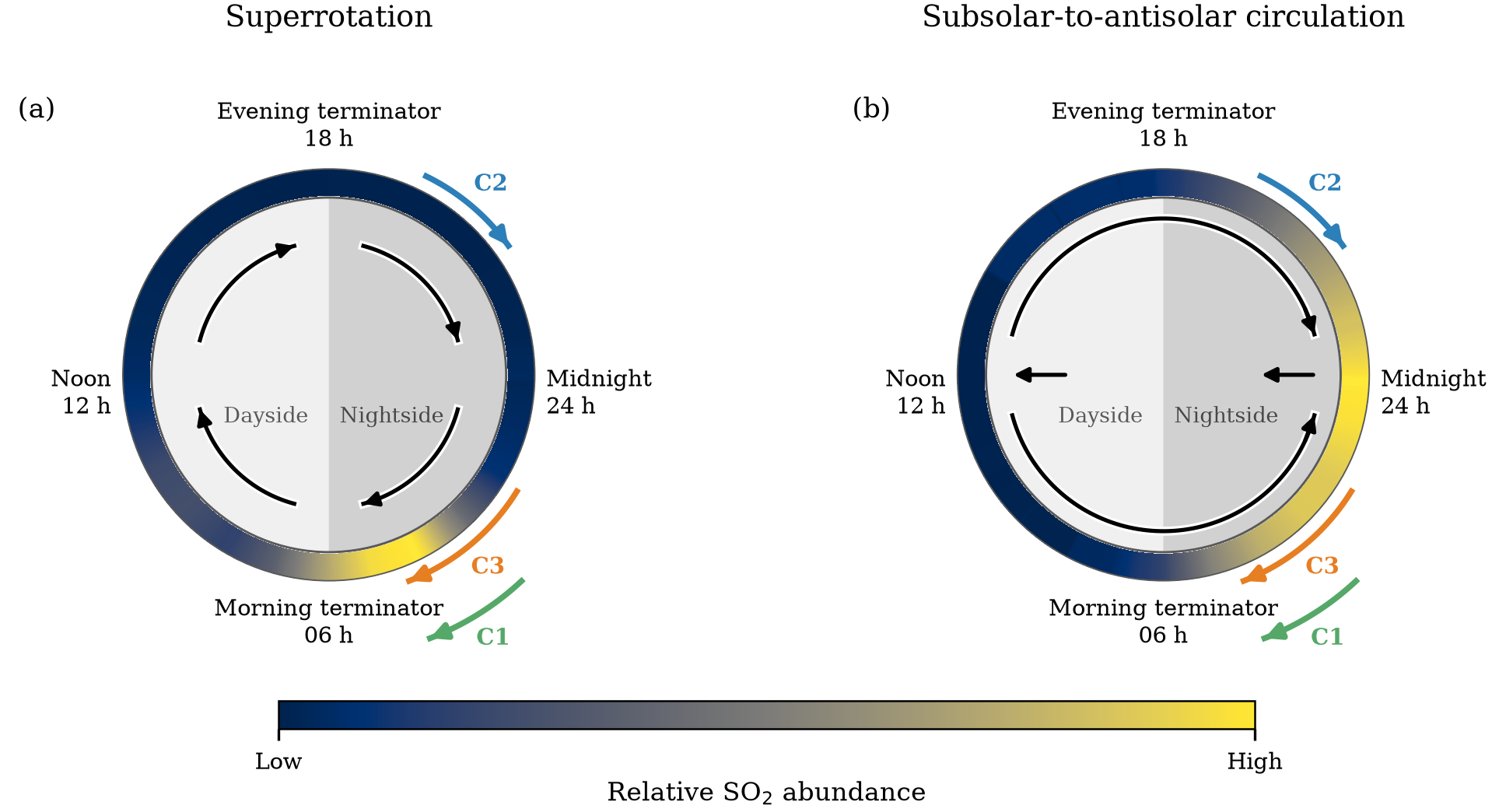}
\caption{
Conceptual illustration of the two dynamical mechanisms that modulate the SO$_2$ LST distribution. Colours around each atmospheric ring qualitatively follow the corresponding SO$_2$ LST profiles of \citet{shao2022local}. Each panel is independently normalised and no absolute abundance values are implied. Black arrows indicate the direction of atmospheric circulation. {Green, blue and orange arrows show the progression of the sub-observer point from the first to the last observation during Campaigns~1, 2 and 3, respectively.} In the superrotation regime, circulation produces a maximum near the morning terminator (06~h LST) and a minimum near the evening terminator (18~h LST). In the SS--AS regime, transport from the subsolar point towards the antisolar point produces a maximum near midnight and a minimum near noon.}
\label{fig:so2_circulation_schematic}
\end{figure*}

In general, SO$_2$ is expected to exhibit a diurnal contrast driven primarily by photodissociation on the dayside, where solar ultraviolet radiation should lead to rapid chemical loss, in contrast to the nightside, where the absence of photodissociation allows SO$_2$ to accumulate. Such a feature has been observed by \citet{sandor2010sulfur}. This basic day--night asymmetry is likely to be further modulated by global atmospheric dynamics, giving rise to a diurnal cycle. Our probed altitude range covers the transition between superrotation and the SS--AS circulation, which modulate the diurnal cycle in distinct ways. In the superrotation-dominated regime, upward transport from more SO$_2$-rich layers, likely driven by thermal tides and diffusion \citep{zasova2007structure}, could enrich SO$_2$ on the nightside. This gas would then be transported to the dayside by superrotation, where it would be efficiently photodissociated and contribute to the formation of H$_2$SO$_4$ aerosols that return to deeper atmospheric layers. This circulation pattern is expected to lead to a maximum in SO$_2$ abundance near the morning terminator (06~h LST) and a minimum near the evening terminator (18~h LST), corresponding to the longest effective timescales for nightside accumulation and dayside loss, respectively. In the SS--AS-dominated regime, SO$_2$ is upwelled near the subsolar point, transported towards the antisolar point, and subsequently subducted back to deeper layers within the superrotation-dominated regime, where the gas is transported back to the dayside. This is expected to produce a nightside maximum in SO$_2$ abundance near midnight (24~h LST) and a minimum near midday (12~h LST). This is supported by \citet{belyaev2012vertical}, who find abundances at about 95 km to be 3--4 times higher at midnight than at the terminators. A conceptual illustration of these two dynamical regimes is shown in Fig.~\ref{fig:so2_circulation_schematic}.

H$_2$O, by contrast, should not exhibit a significant LST dependence, given its much longer photodissociation timescale compared to the zonal transport timescale \citep{mahieux2024unexpected}. None of the observations support the presence of an LST dependence for H$_2$O, although the existing data are either too sparse \citep{sandor2005water,gurwell2007swas,piccialli2017mapping} or, in the case of occultation measurements, largely limited to the terminators and therefore poorly suited to addressing this question \citep{bertaux2007warm,fedorova2008hdo,chamberlain2020soir}.

Based on the above understanding, we attribute the SO$_2$ abundance differences between Campaigns~2 and 3 to superrotation. In the superrotation-dominated scenario, Campaign~2 predominantly samples the first half of the nightside accumulation region, where the gas has only recently rotated from the dayside into the nightside, together with adjacent dayside regions in which SO$_2$ has already been strongly depleted by photodissociation. By contrast, Campaign~3 probes the second half of the nightside accumulation region, along with neighbouring dayside regions where photodissociation has only recently commenced and relatively high SO$_2$ abundances are still present. A purely SS--AS-dominated scenario would be expected to produce a more symmetric distribution between the two halves of the nightside. {It should be noted that this interpretation assumes a nearly constant SO$_2$ inventory in the upper mesosphere between Campaigns~2 and~3, which were close in time but still several weeks apart. Therefore, alternative explanations for the SO$_2$ increase between the two campaigns cannot be ruled out, such as a coincident dynamical event.}

The campaign-wide increasing and decreasing trends in SO$_2$ observed during Campaigns~2 and~3 can be understood in terms of the evolving LST coverage. In both campaigns, as the sampled LST range progressively narrows towards midnight, the peak beam response also shifts towards midnight, and the observed SO$_2$ abundance increases. This behaviour is consistent with a nightside maximum near midnight, indicating that the SS–AS circulation contributes in addition to superrotation. The combined influence may to some extent account for the less monotonic behaviour observed in Campaign~3 compared with Campaign~2. Specifically, during the earlier observing dates of Campaign~3, the sampled LST range does not yet, or only just, extend into the dayside. Under these conditions, superrotation still acts to increase the observed SO$_2$ abundance and may offset the decrease induced by the SS--AS contribution, thereby producing the broadly stable trend. As the LST coverage extends further into the dayside during later dates, where photochemical loss begins to catch up with nightside accumulation in contributing to the disc-integrated signal, superrotation begins to drive a decrease instead. The combined effects of the two dynamical components could therefore explain the initially stable behaviour followed by the late-campaign decline observed in Campaign~3.


\subsection{LST model}

An independent LST dependence study was conducted by \citet{shao2022local}, who predicted SO$_2$ and H$_2$O abundances as a function of LST based on a three-dimensional general circulation model (GCM) combined with a two-dimensional chemical transport model (CTM). Their model probes altitudes of 84.8 and 94.8~km, which lie within our sensitivity range. In their model, the circulation at 84.8~km is dominated by superrotation, and at 94.8~km the SS--AS circulation is superimposed on the superrotation.

\begin{figure*}
\centering
\includegraphics[width=\textwidth]{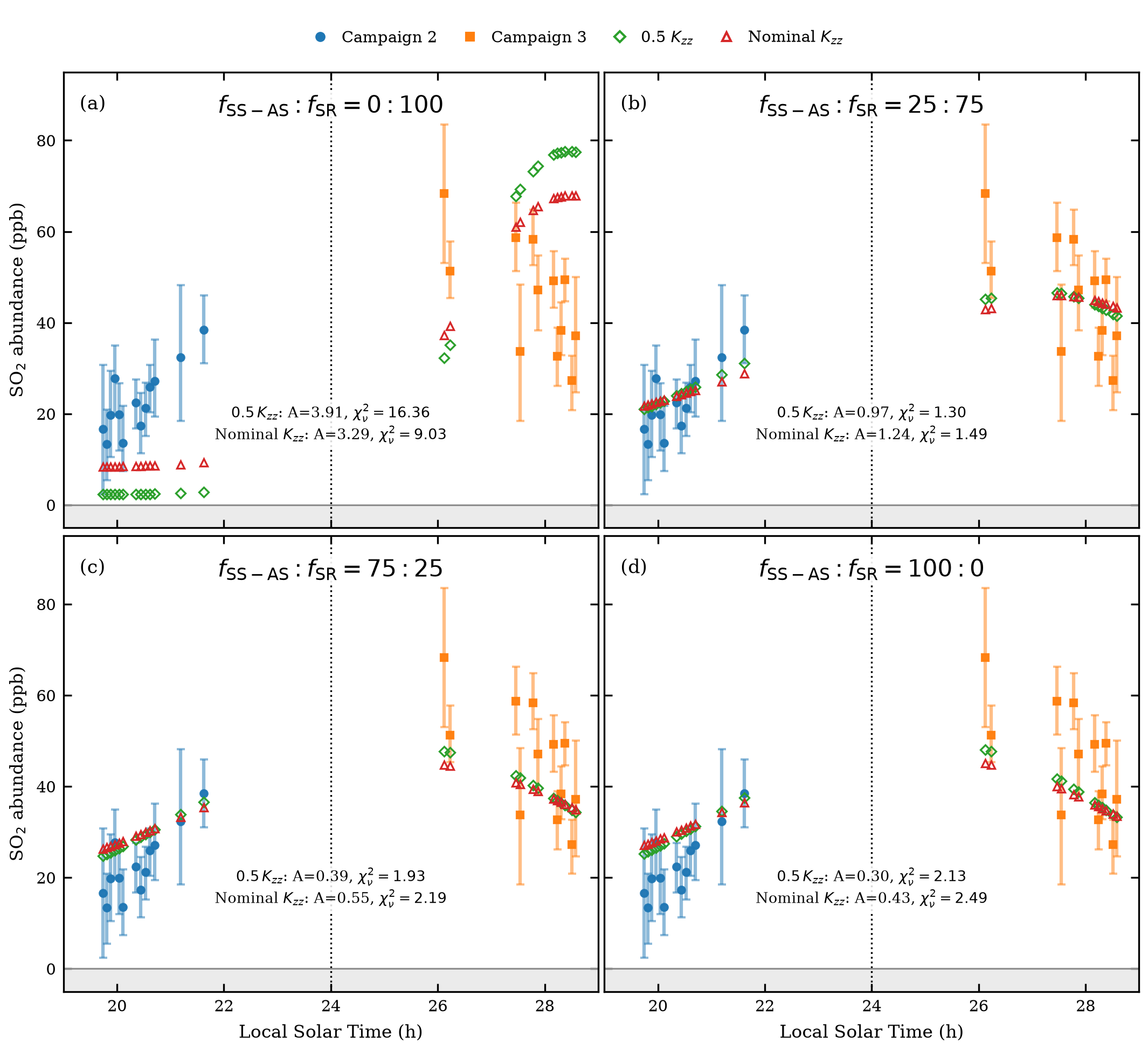}
\caption{
Comparison of the observed SO$_2$ abundances with the model predictions of \citet{shao2022local} for different relative weightings of the SS--AS-dominated model at 94.8~km and the superrotation-dominated model at 84.8~km. All panels show the same observational dataset and compare the nominal and half-nominal $K_{zz}$ profiles. For each model, $A$ denotes the multiplicative scaling factor applied so that the mean modelled abundance matches the observed mean, while $\chi_\nu^2$ denotes the reduced chi-squared value of the comparison.
}
\label{model}
\end{figure*}

For each observing date, the model LST profile \(X_{\rm mod}(T)\) is converted to a disc-integrated abundance using the projected Venus disc and the JCMT beam response, {with the corresponding integration given in Appendix~\ref{app:disc_integration}.} The comparison for SO$_2$ is shown in Fig.~\ref{model}. We explore different relative weightings of the model profiles at two altitudes, each dominated by a distinct dynamical mechanism, using both the nominal and half-nominal vertical diffusion coefficient ($K_{zz}$) profiles implemented by \citet{shao2022local}. The modelled values are scaled so that their mean matches that of the observations, since our focus is on relative variability rather than the absolute abundance scale. The model predictions agree much better with the observations when both mechanisms are included. In particular, a purely superrotation-dominated scenario fails to reproduce the campaign-wide trends, whereas a purely SS--AS-dominated scenario does not produce a sufficiently large difference between the two campaigns.

\begin{figure}
\centering
\includegraphics[width=0.92\columnwidth]{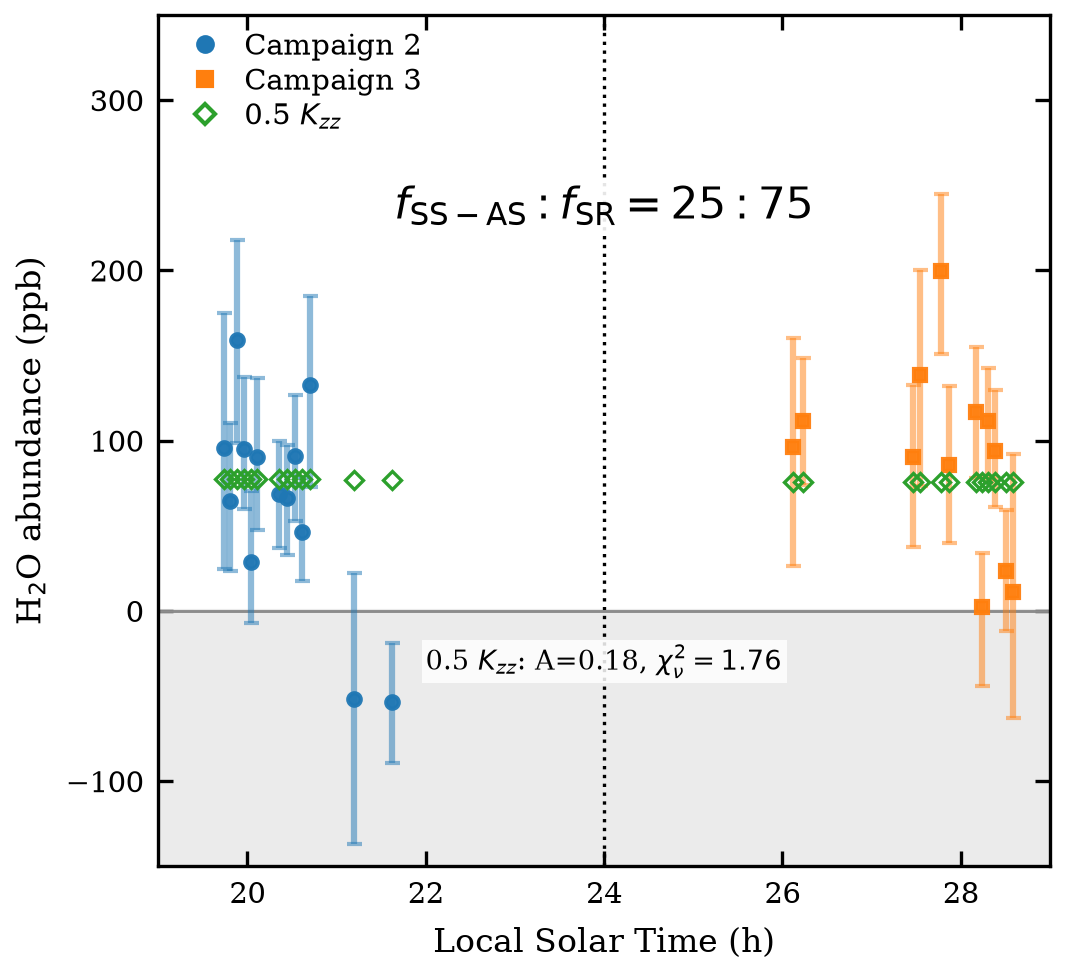}
\caption{
Comparison for H$_2$O, same as Fig.~\ref{model}. The parameters adopt the best-fitting configuration derived for SO$_2$.
}
\label{model2}
\end{figure}

\begin{figure}
\centering
\vspace{0.2em}

    \includegraphics[width=0.90\columnwidth]{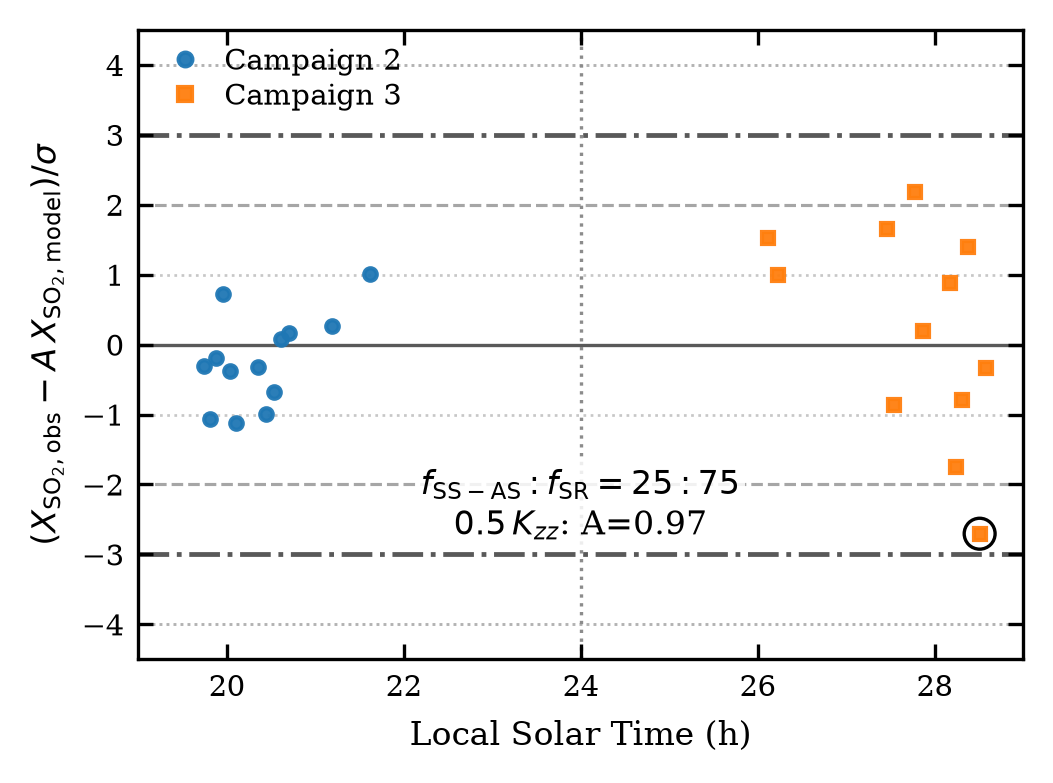}\\[-0.4em]
    
    \includegraphics[width=0.90\columnwidth]{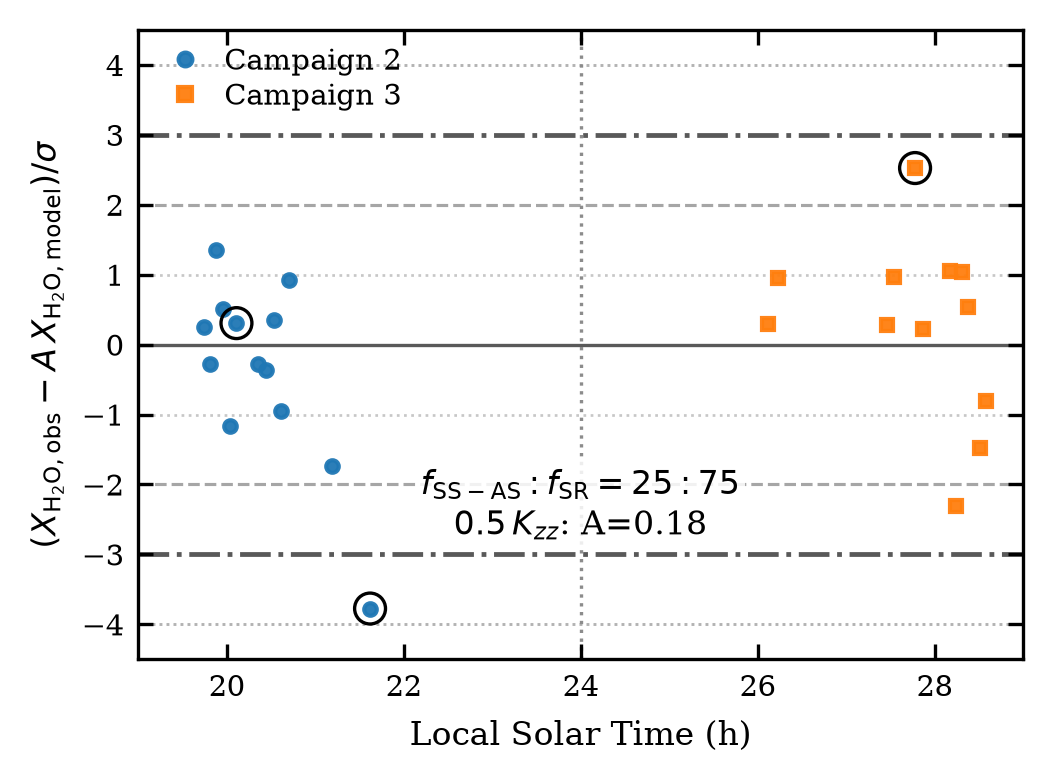}
    
    \caption{
    Top: Deviation significance of the observed SO$_2$ abundance relative to the model prediction, 
    expressed as $(X_{\rm obs}-AX_{\rm mod})/\sigma$, for the 0.5\,K$_{zz}$ model with 
    25\% SS--AS and 75\% superrotation weighting. 
    Bottom: Same as above but for H$_2$O. The circled points mark the excursions identified in Section~\ref{dtd}.
    }
    
    \label{dev}
\end{figure}

Among the four tested configurations, the best agreement is obtained for a weighting of 25\% SS--AS-dominated model and 75\% superrotation-dominated model, in terms of both the absolute abundance level and the campaign-wide trends. In addition, the $0.5\times K_{zz}$ profile provides a better match to the observations than the nominal $K_{zz}$ profile, as it produces steeper gradients that more closely resemble the observed behaviour. Although a strict determination of the $K_{zz}$ profile and the dynamical processes in our probed region is beyond the scope of this study, our results demonstrate strong potential for constraining them.

Using the best-fitting configuration for SO$_2$, we conduct an analogous comparison for H$_2$O, shown in Fig.~\ref{model2}. An exploration of multiple configurations, as performed for SO$_2$, is not carried out for H$_2$O, as varying the relative weighting of the two mechanisms or the $K_{zz}$ profile produces only negligible changes in the outputs. Our observations are generally consistent with the model, showing minimal LST dependence for H$_2$O despite a few days potentially exhibiting short-term excursions. The model predicts a higher overall abundance than the observed average, which is not unexpected, since the absolute scale depends strongly on the lower boundary condition, set to 1 ppm at 58 km in the model. Moreover, the model values correspond to only two discrete altitudes, whereas our measurements represent a weighted contribution over a broad vertical range.

We further evaluate the agreement between the observations and the model predictions by computing the deviation in units of the observational uncertainty, defined as $(X_{\rm obs} - AX_{\rm mod})/\sigma_{\rm obs}$, where $A$ is the scaling factor applied to the model to match the observed mean abundance. The results are shown in Fig.~\ref{dev}. No model prediction deviates from the observations by more than the 3-$\sigma$ level, except for H$_2$O on 27 July in Campaign~2. The date 27 July is one of the excursions identified in Section~\ref{dtd}, with the others also highlighted in the plot. This sudden depletion suggests the presence of short-term variation on a timescale of up to nine days. This comparison is not intended to imply that the model perfectly reproduces all observations, or that all residual variability is purely systematic. Rather, the purpose of this comparison is to demonstrate that the LST model {can explain} the primary behaviour of the two species in the absence of a long-term effect, while the remaining short-term fluctuations are moderate and only rarely drive the abundances to statistically significant deviations from the model predictions.

{A concern is that the LST-dependent trend in SO$_2$ could be attributed to interday fluctuations in the temperature profiles. We examine this possibility by perturbing the daily temperature profiles using the VCD estimates of temporal temperature variability and repeating the retrievals for these perturbed profiles. We then assess the agreement with the LST model across 10,000 random combinations of the resulting daily abundances. The results indicate that the agreement with the LST model is still valid under the tested temperature perturbations, as discussed in Appendix~\ref{app:vcd_variability}.}

\section{Conclusions}

We have presented three multi-week, disc-integrated JCMT campaigns monitoring H$_2$O and SO$_2$ on Venus at altitudes of approximately 80--110~km. The extended temporal coverage and dense sampling provide the longest continuous ground-based dataset for exploring Venus’ upper mesosphere, yielding important insights into the long-term, short-term, and LST-dependent variability of these two species.

We propose that the long-term variability of SO$_2$ in the upper mesosphere may be coupled to the long-term trend at the cloud tops through vertical supply. This further suggests that the potential reservoir enriching SO$_2$ in the upper mesosphere is also linked to lower altitudes. This places constraints on its origin, making the oxidation of polysulphur derived from OCS less favoured than the photodissociation of H$_2$SO$_4$ vapour \citep{zhang2010photolysis,zhang2012sulfur}. Dynamical processes remain a possible explanation for the SO$_2$ inversion \citep{stolzenbach2015three}. Long-term variability in H$_2$O is verified here for the first time, and is interpreted as a response to the long-term change in SO$_2$ through the formation of H$_2$SO$_4$.

We find that SO$_2$ {variability is consistent with an LST-dependent trend}, which is well reproduced by the model of \citet{shao2022local}. The persistent agreement between the model predictions and the observations over two multi-week campaigns indicates that SO$_2$ variability in the upper mesosphere is {probably} governed primarily by deterministic photochemical--dynamical processes. Although short-term variability is likely to be present, it does not dominate the overall abundance structure. We therefore suggest that any interpretation of mesospheric SO$_2$ variability should explicitly account for LST dependence.

Comparison of our observations with the model further demonstrates the potential to constrain atmospheric parameters such as $K_{zz}$ and the relative contributions of dynamical mechanisms within the probed layer. In our analysis, the observed trends are best reproduced when both superrotation and SS--AS circulation are included, with reduced vertical diffusion ($0.5\times K_{zz}$) providing improved agreement.

H$_2$O shows no statistically significant evidence of LST dependence, consistent with model expectations and previous observations. Its abundance remains broadly stable within each campaign, indicating overall stability, in agreement with \citet{bertaux2007warm} and \citet{fedorova2008hdo}. At the same time, we identify short-term depletion events on timescales of up to nine days, consistent with the variability reported by \citet{sandor2005water}, \citet{gurwell2007swas}, and \citet{chamberlain2020soir}.

A central component of this study is the development of a consistent pipeline to mitigate standing-wave systematics. The same Fourier-based ripple removal, forward-modelling retrieval, and injection-based error quantification are applied uniformly to all species and campaigns. This provides a reliable reference for future JCMT observations of Venus and for other datasets affected by standing-wave contamination. We also suggest revisiting earlier submm/mm datasets suffering from the same issue using the pipeline developed in this work \citep[e.g.][]{sandor2005water,sandor2010sulfur,gurwell2007swas}.

Future work will present a sulphur-budget study based on observations of multiple sulphur-bearing species across the three campaigns, including OCS, SO, and H$_2$SO$_4$ \citep{Tang2026}. Because sulphur chemistry largely redistributes material among molecular reservoirs while approximately conserving the total sulphur budget, simultaneous constraints on multiple sulphur species will help to determine whether occasional fluctuations arise from chemical partitioning or dynamical transport. In this way, sulphur-bearing molecules can serve as tracers of the physical and chemical processes operating in the upper mesosphere, providing insight into outstanding questions such as the origins of the SO$_2$ inversion profile and short-term variability.

Two further JCMT--Venus campaigns have been undertaken in April 2025 and July--August 2026. Campaign~4 sampled Venus near the maximum of Solar Cycle~25, while Campaign~5 provides coverage of predominantly dayside LSTs. These observations extend the temporal baseline and will provide additional information on the long-term, interday, and LST-dependent variability of these species.

\section*{Acknowledgements}

{We thank the referee for recommending the Venus Climate Database.} These observations were obtained with the James Clerk Maxwell Telescope, which is operated by the East Asian Observatory on behalf of the National Astronomical Observatory of Japan, the Academia Sinica Institute of Astronomy and Astrophysics, the Korea Astronomy and Space Science Institute, the National Astronomical Research Institute of Thailand, and the Center for Astronomical Mega-Science, as well as through the National Key R\&D Program of China (No.~2017YFA0402700). Additional funding support is provided by the Science and Technology Facilities Council of the United Kingdom and by participating universities and organisations in the United Kingdom and Canada. N\={a}makanui was constructed and funded by ASIAA in Taiwan, with funding for the mixers provided by ASIAA and, at 230~GHz, by EAO. The N\={a}makanui instrument is a backup receiver for the GLT. This work has made use of the Starlink Software Collection \citep{currie2014starlink} and is supported in part by the Fundamental Fund of Thailand Science Research and Innovation (TSRI) through the National Astronomical Research Institute of Thailand (Public Organization) (FFB690078/0269).
\section*{Data Availability}

The code used in this study is available from Wei Tang upon request, and the data are available from the JCMT archive at \url{https://www.cadc-ccda.hia-iha.nrc-cnrc.gc.ca}.




\bibliographystyle{mnras}
\begingroup\renewcommand{\href}[2]{#2}\bibliography{example}\endgroup



\appendix
\raggedbottom

\clearpage
\twocolumn[{\begin{minipage}{\textwidth}
\begingroup

\section{Sensitivity to the assumed temperature profile}

\label{app:temperature}

Figure~\ref{fig:temperature_comparison} shows absolute and relative differences between the daily abundances derived using the adopted VCD temperature profiles and the Venus International Reference Atmosphere (VIRA) \citep{seiff1985models}. The maximum absolute differences are 12.9~ppb for H$_2$O and 2.2~ppb for SO$_2$. Excluding SO$_2$ in Campaign~1 and H$_2$O on 19 September 2023, where near-zero abundances make relative differences sensitive to small absolute changes, the maximum relative differences are 3.4\% for SO$_2$ and 7.7\% for H$_2$O.

\endgroup

\par\medskip\noindent\begin{minipage}{\textwidth}
\centering
\includegraphics[width=0.95\textwidth]{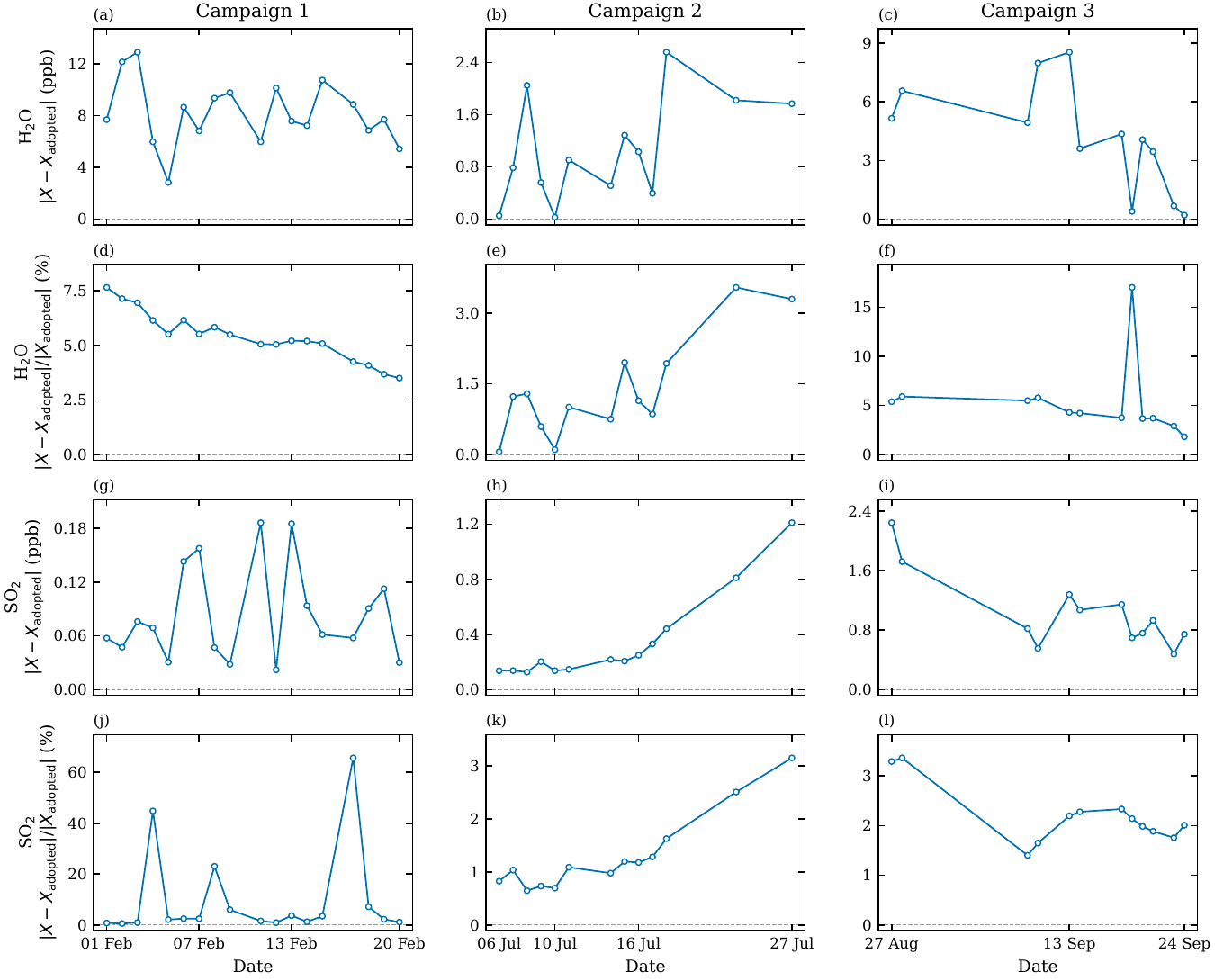}
\captionsetup{hypcap=false}
\captionof{figure}{{Daily abundance differences derived using VIRA relative to the adopted temperature profiles. The first and third rows show the absolute differences $|X-X_{\rm adopted}|$, while the second and fourth rows show the relative differences $|X-X_{\rm adopted}|/|X_{\rm adopted}|$, for H$_2$O and SO$_2$, respectively. Columns correspond to Campaigns~1--3. Relative differences become sensitive to small absolute changes when $X_{\rm adopted}$ is close to zero.}}
\label{fig:temperature_comparison}
\end{minipage}

\end{minipage}}]

\clearpage
\twocolumn[{\begin{minipage}{\textwidth}
\begingroup

\section{Sensitivity to the assumed D/H profile}
\label{app:dh}

We assess the sensitivity of the H$_2$O abundances to the assumed D/H profile by replacing the mean profile of \citet{mahieux2024unexpected} with a constant value of $200\times$VSMOW or the lower or upper envelope of \citet{mahieux2024unexpected}. We find that changing the D/H profile rescales the derived daily abundances by an approximately constant factor. These factors vary by less than 6\% between dates within each campaign (Fig.~\ref{fig:dh_abundance_factor}). They have negligible influence on the relative temporal behaviour, provided that the same D/H profile is applied to all dates. The anomalous point in Campaign 3 corresponds to 19 September, when the adopted abundance was only about 2~ppb, making the ratio sensitive to small absolute differences.
\endgroup

\par\medskip\noindent\begin{minipage}{\textwidth}
\centering
\includegraphics[width=0.95\textwidth]{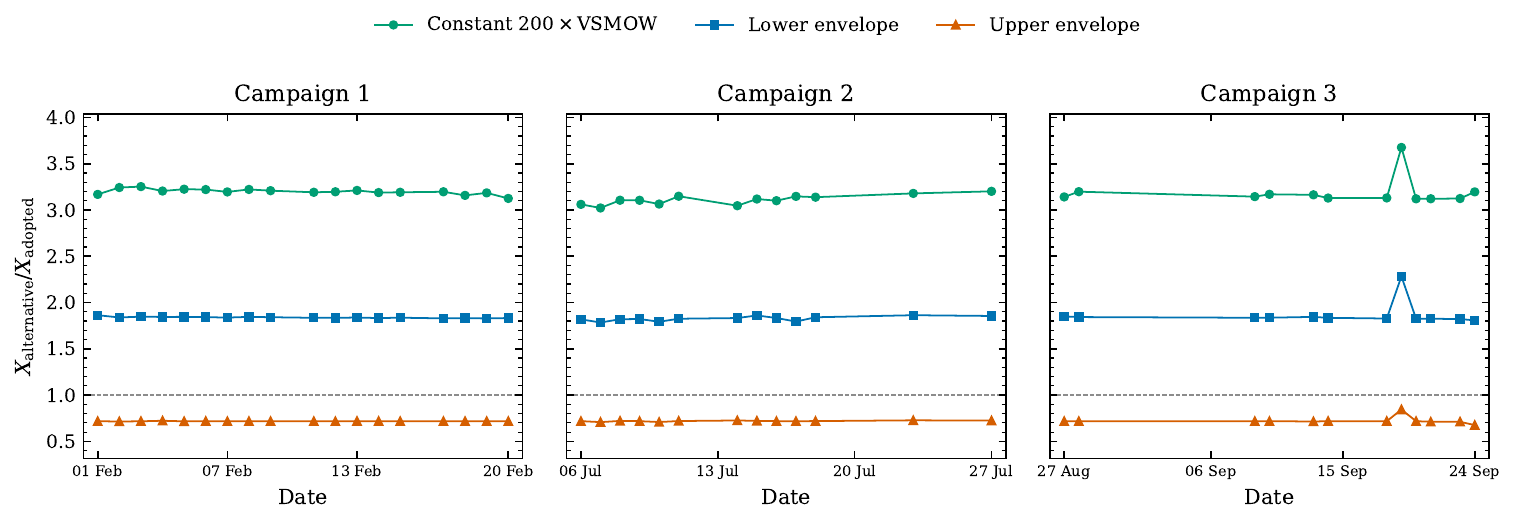}
\captionsetup{hypcap=false}\captionof{figure}{{Daily H$_2$O abundance ratios, $X_{\rm alternative}/X_{\rm adopted}$. The three panels show Campaigns~1--3. Green circles denote a constant D/H of $200\times$VSMOW, while blue squares and orange triangles denote the lower and upper envelopes of \citet{mahieux2024unexpected}, respectively. The dashed line marks unity.}}
\label{fig:dh_abundance_factor}
\end{minipage}

\end{minipage}}]

\clearpage
\twocolumn[{\begin{minipage}{\textwidth}

\begingroup

\section{Daily abundances and observing geometry}
\label{app:daily_data}
\endgroup

\par\medskip\noindent\begin{minipage}{\textwidth}
\centering
\small
\captionsetup{hypcap=false}\captionof{table}{Daily H$_2$O and SO$_2$ abundances and observing geometry for the three campaigns. The Venus LST range subtended by the JCMT beam at the equator is computed assuming a spherical-disc geometry, with an effective beam diameter corresponding to the 10\% power level of a Gaussian beam ($\theta_{\rm beam,10\%}=33.54~\mathrm{arcsec}$ for $\theta_{\rm FWHM}=18.4~\mathrm{arcsec}$).}
\label{abundance}
\begin{tabular}{lcccccccc}
\hline
Date &
Illum. &
Angular diameter &
Disc-centre LST &
LST range &
H$_2$O abundance &
H$_2$O error &
SO$_2$ abundance &
SO$_2$ error \\
 & (\%) & (arcsec) & (h) & (h) & (ppb) & $+/-$ & (ppb) & $+/-$ \\
\hline
\multicolumn{9}{l}{\textit{Campaign 1 (Feb 2022)}} \\
01~Feb & 15.9 & 48.6 & 3.13 & 0.22--6.04 & 100 & $+50/-52$ & 8  & $+5/-5$ \\
02~Feb & 16.8 & 47.8 & 3.23 & 0.26--6.20 & 170 & $+83/-82$ & 9  & $+13/-11$ \\
03~Feb & 17.7 & 47.0 & 3.32 & 0.29--6.35 & 186 & $+52/-49$ & 8   & $+6/-6$ \\
04~Feb & 18.6 & 46.3 & 3.41 & 0.32--6.50 & 97  & $+36/-39$ & 0   & $+7/-6$ \\
05~Feb & 19.5 & 45.5 & 3.49 & 0.32--6.66 & 52  & $+47/-46$ & 1   & $+10/-8$ \\
06~Feb & 20.4 & 44.8 & 3.58 & 0.35--6.81 & 141 & $+49/-59$ & -6  & $+13/-10$ \\
07~Feb & 21.3 & 44.0 & 3.66 & 0.35--6.97 & 124 & $+34/-40$ & -6  & $+8/-7$ \\
08~Feb & 22.2 & 43.3 & 3.75 & 0.37--7.13 & 160 & $+42/-38$ & 0   & $+7/-6$ \\
09~Feb & 23.1 & 42.6 & 3.83 & 0.37--7.29 & 178 & $+45/-49$ & 0   & $+6/-7$ \\
11~Feb & 24.9 & 41.2 & 3.99 & 0.36--7.62 & 118  & $+36/-40$ & 12   & $+6/-7$ \\
12~Feb & 25.7 & 40.5 & 4.06 & 0.33--7.79 & 201 & $+34/-37$ & 2  & $+6/-7$ \\
13~Feb & 26.6 & 39.9 & 4.14 & 0.33--7.95 & 145 & $+36/-40$ & -5  & $+7/-6$ \\
14~Feb & 27.4 & 39.2 & 4.21 & 0.29--8.13 & 139 & $+29/-31$ & -8 & $+5/-5$ \\
15~Feb & 28.2 & 38.6 & 4.28 & 0.26--8.30 & 211 & $+46/-38$ & -2  & $+6/-6$ \\
17~Feb & 29.8 & 37.4 & 4.41 & 0.16--8.66 & 208 & $+43/-40$ & 0  & $+6/-6$ \\
18~Feb & 30.6 & 36.8 & 4.48 & 0.10--8.86 & 168 & $+33/-33$ & -1  & $+4/-6$ \\
19~Feb & 31.4 & 36.2 & 4.54 & 0.01--9.07 & 210 & $+31/-26$ & -5  & $+5/-4$ \\
20~Feb & 32.1 & 35.7 & 4.60 & 23.94--9.26 & 155 & $+37/-36$ & 3  & $+5/-7$ \\

\hline
\multicolumn{9}{l}{\textit{Campaign 2 (Jul 2023)}} \\
06~Jul & 28.0 & 36.2 & 19.74 & 15.21--0.27 & 96  & $+80/-71$ & 17 & $+14/-14$ \\
07~Jul & 27.2 & 36.8 & 19.81 & 15.43--0.19 & 64 & $+46/-41$ & 13  & $+7/-8$ \\
08~Jul & 26.4 & 37.4 & 19.88 & 15.63--0.13 & 159 & $+58/-61$ & 20 & $+10/-9$ \\
09~Jul & 25.5 & 38.0 & 19.96 & 15.83--0.09 & 95  & $+42/-35$ & 28 & $+7/-8$ \\
10~Jul & 24.6 & 38.6 & 20.04 & 16.02--0.06 & 29  & $+42/-36$ & 20 & $+7/-8$ \\
11~Jul & 23.8 & 39.2 & 20.11 & 16.19--0.03 & 90 & $+46/-43$ & 14 & $+8/-6$ \\
14~Jul & 21.1 & 41.2 & 20.35 & 16.72--23.98 & 69  & $+31/-32$ & 22 & $+5/-6$ \\
15~Jul & 20.2 & 41.9 & 20.44 & 16.90--23.98 & 66  & $+31/-33$ & 17 & $+7/-6$ \\
16~Jul & 19.3 & 42.5 & 20.53 & 17.06--0.00 & 91 & $+36/-38$ & 21 & $+6/-6$ \\
17~Jul & 18.4 & 43.2 & 20.61 & 17.22--0.00 & 46  & $+32/-29$ & 26 & $+5/-6$ \\
18~Jul & 17.5 & 43.9 & 20.70 & 17.38--0.02 & 133 & $+52/-60$ & 27 & $+9/-8$ \\
23~Jul & 12.9 & 47.5 & 21.19 & 18.20--0.18 & -51 & $+74/-86$ & 32 & $+16/-14$ \\
27~Jul &  9.4 & 50.3 & 21.62 & 18.83--0.41 & -54 & $+35/-36$ & 38 & $+7/-7$ \\

\hline
\multicolumn{9}{l}{\textit{Campaign 3 (Aug--Sep 2023)}} \\
27~Aug &  7.5 & 52.7 & 2.12 & 23.49--4.75 & 96 & $+64/-69$ & 68 & $+15/-15$ \\
28~Aug &  8.3 & 52.1 & 2.23 & 23.56--4.90 & 111 & $+37/-37$ & 51 & $+6/-6$ \\
09~Sep & 19.1 & 43.8 & 3.46 & 0.13--6.79 & 90 & $+43/-52$ & 59 & $+8/-7$ \\
10~Sep & 20.0 & 43.2 & 3.54 & 0.15--6.93 & 138 & $+62/-65$ & 34 & $+15/-15$ \\
13~Sep & 22.6 & 41.2 & 3.78 & 0.15--7.41 & 200 & $+45/-49$ & 58 & $+6/-6$ \\
14~Sep & 23.5 & 40.6 & 3.87 & 0.16--7.58 & 86  & $+46/-46$ & 47 & $+8/-9$ \\
18~Sep & 26.9 & 38.2 & 4.17 & 0.08--8.26 & 117 & $+38/-39$ & 49 & $+6/-6$ \\
19~Sep & 27.8 & 37.6 & 4.24 & 0.03--8.45 & 2  & $+32/-46$ & 33 & $+6/-6$ \\
20~Sep & 28.6 & 37.0 & 4.31 & 23.98--8.64 & 111 & $+32/-34$ & 38 & $+6/-6$ \\
21~Sep & 29.4 & 36.5 & 4.38 & 23.93--8.83 & 94  & $+36/-33$ & 49 & $+5/-5$ \\
23~Sep & 31.0 & 35.4 & 4.51 & 23.76--9.26 & 23   & $+36/-35$ & 27 & $+5/-6$ \\
24~Sep & 31.8 & 34.9 & 4.58 & 23.65--9.51 & 11 & $+81/-74$ & 37 & $+13/-12$ \\
\hline
\end{tabular}
\end{minipage}

\end{minipage}}]

\clearpage
\section{Campaign-wide trend assessment}
\label{mt}
To test whether campaign-wide monotonic trends are present and are not driven by systematic effects, we apply Kendall's rank-correlation test. We compare the trend measured from the observed abundances with those obtained from 102 line-free channels. For SO$_2$ in Campaign~2, Kendall's $\tau$ for the observed abundances is 0.60 ($p=0.005$), compared with a maximum positive $\tau$ of 0.59 from the line-free channels. This supports the presence of a monotonic trend in the observed abundances. None of the other cases pass this test (Fig.~\ref{fig:kendall_trend}).

\par\medskip\noindent\begin{minipage}{\columnwidth}
\centering
\includegraphics[width=\columnwidth]{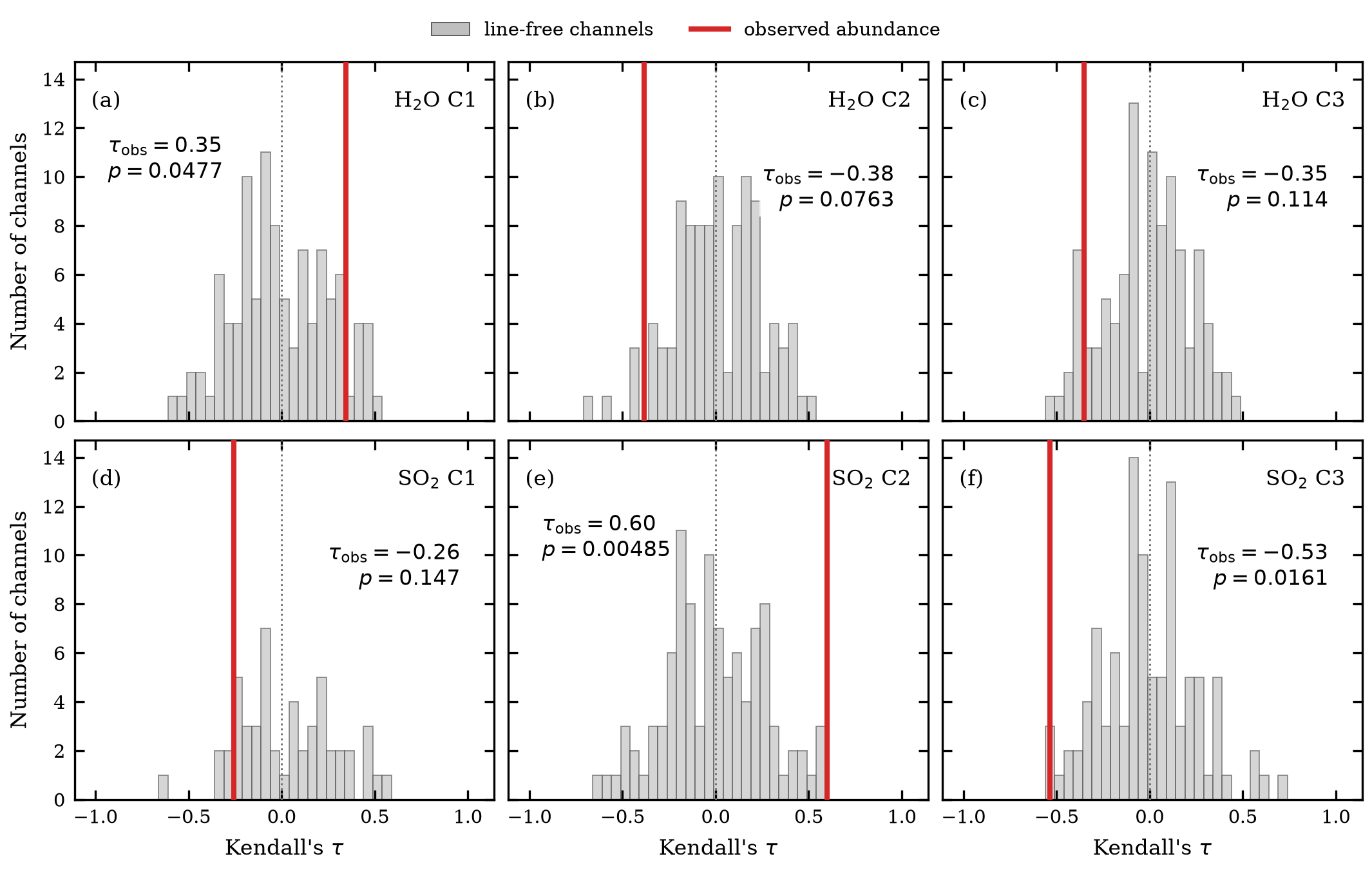}
\captionsetup{hypcap=false}
\captionof{figure}{
Kendall rank-correlation coefficients for the observed abundances compared with those derived from the line-free channels. Grey histograms show the distribution of Kendall's $\tau$ values obtained from the 102 line-free channels, while the red vertical line marks the value measured from the observed daily abundances. Only SO$_2$ in Campaign~2 shows an observed monotonic trend that exceeds the maximum trend found in the line-free channels.
}
\label{fig:kendall_trend}
\end{minipage}


\clearpage
\begingroup

\section{Disc integration}
\label{app:disc_integration}

For each observing date, the model LST profile \(X_{\rm mod}(T)\) is converted to a disc-integrated abundance using the projected Venus disc and the JCMT beam response. Venus is represented as a sphere of angular radius \(R\), and the beam is approximated as a circular Gaussian with \(\theta_{\rm FWHM}=18.4~\mathrm{arcsec}\). An LST offset \(\Delta T\) from the disc centre corresponds to a longitude offset
\begin{equation}
\phi=\frac{\pi}{12}\Delta T .
\end{equation}
For an element at latitude \(\lambda\) and longitude offset \(\phi\), the coordinates on the projected plane are
\begin{equation}
x(\lambda,\phi)=R\cos\lambda\sin\phi
\end{equation}
and
\begin{equation}
y(\lambda)=R\sin\lambda .
\end{equation}
The JCMT beam response at this position is
\begin{equation}
K_{\rm beam}(\lambda,\phi)=
\exp\!\left[
-4\ln2\,
\frac{
R^2\left(\cos^2\lambda\sin^2\phi+\sin^2\lambda\right)
}{
\theta_{\rm FWHM}^2
}
\right].
\end{equation}
The projection factor is
\begin{equation}
K_{\rm proj}(\lambda,\phi)
=\cos\lambda\cos\phi .
\end{equation}
Multiplying this by the surface-area element
\begin{equation}
{\rm d}A=R^2\cos\lambda\,{\rm d}\lambda\,{\rm d}\phi
\end{equation}
gives the corresponding projected-area element
\begin{equation}
{\rm d}A_{\rm proj}
=K_{\rm proj}(\lambda,\phi)\,{\rm d}A
=R^2\cos^2\lambda\cos\phi\,
{\rm d}\lambda\,{\rm d}\phi .
\end{equation}
Since \(X_{\rm mod}\) depends only on LST and is assumed to be independent of latitude, the total weight associated with a given longitude offset is obtained by integrating over latitude:
\begin{equation}
K(\phi)=
R^2\cos\phi
\int_{-\pi/2}^{+\pi/2}
K_{\rm beam}(\lambda,\phi)
\cos^2\lambda\,{\rm d}\lambda .
\end{equation}
The abundance corresponding to an observation centred at disc-centre LST \(T_0\) is then
\begin{equation}
\langle X\rangle_{\rm day} =
\frac{\displaystyle\int_{-6\,{\rm h}}^{+6\,{\rm h}}
X_{\rm mod}(T_0+\Delta T)\,
K[\phi(\Delta T)]\,{\rm d}\Delta T}
{\displaystyle\int_{-6\,{\rm h}}^{+6\,{\rm h}}
K[\phi(\Delta T)]\,{\rm d}\Delta T}.
\end{equation}
\endgroup

\clearpage
\twocolumn[{\begin{minipage}{\textwidth}
\begingroup

\section{Sensitivity of the LST comparison to temperature variability}

\label{app:vcd_variability}

We assess the effect of temporal temperature variability on the LST-model comparison using the temporal RMS of the temperature profiles provided by the VCD \citep{lebonnois2010superrotation,lebonnois2016wave,gilli2017thermal,martinez2023variability}. For each of the 25 observing dates in Campaigns~2 and~3, we generate five perturbed temperature profiles by multiplying the VCD temperature RMS by a random coefficient drawn from a standard normal distribution and adding the resulting perturbation to the adopted temperature profile.

We repeat the same abundance retrieval and uncertainty estimation procedures for each daily perturbed case. We then generate 10\,000 random combinations by independently selecting one of the five cases for each observing date. For each combination, we repeat the comparison with the 25:75 SS--AS:superrotation model and half-nominal $K_{zz}$ \citep{shao2022local}, adjusting $A$ to match the mean abundance. The resulting distributions of reduced $\chi^2_\nu$ are shown in Figure~\ref{fig:vcd_temperature_chi}. The median $\chi^2_\nu$ values are 1.73 for H$_2$O and 1.34 for SO$_2$, with 16th--84th percentile ranges of 1.65--1.82 and 1.14--1.56, respectively. The agreement with the LST model is therefore valid under the temperature perturbations considered here.
\endgroup

\par\medskip\noindent\begin{minipage}{\textwidth}
\centering
\includegraphics[width=0.95\textwidth]{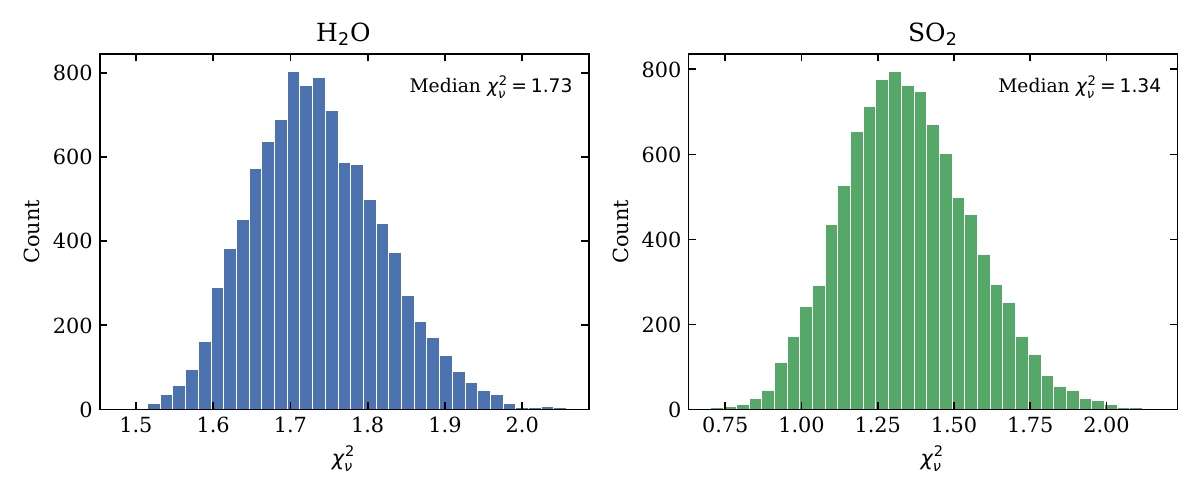}
\captionsetup{hypcap=false}
\captionof{figure}{{Distributions of reduced $\chi^2_\nu$ for the H$_2$O (left) and SO$_2$ (right) LST-model comparisons after perturbing the temperature profiles using the VCD temperature RMS. Each histogram contains 10\,000 random combinations of daily perturbed profiles for Campaigns~2 and~3.}}
\label{fig:vcd_temperature_chi}
\end{minipage}

\end{minipage}}]


\bsp	
\label{lastpage}
\end{document}